\documentclass[12pt]{article}
\usepackage{amsfonts}
\usepackage{amsmath}
\usepackage{amssymb}
\usepackage{graphicx}
\usepackage{color}
\usepackage[all, knot]{xy}
\usepackage{tikz}

\usepackage{ulem}
\usepackage{hyperref}

\usepackage[utf8]{inputenc}
\usepackage{epstopdf}
\usepackage[footnotesize]{caption}
\usepackage{amsthm}
\usepackage{enumitem}
\usepackage{mathrsfs}
\usepackage{mathtools}

\usepackage[margin=3cm]{geometry}

\def \be {\begin{equation}}
\def \ee {\end{equation}}
\def \bea {\begin{eqnarray}}
\def \eea {\end{eqnarray}}
\def \nn {\nonumber}

\def \rr {\raise.35ex\hbox{\small $\prime$}\kern-.17em{\mbox{\large $\imath$}}}

\def \dels {\partial\kern-.6em /\kern.1em}
\def \As {{A\kern-.5em / \kern.5em}}
\def \Ds {D\kern-.7em / \kern.5em}

\def \ks {k\kern-.5em /}
\def \ls {l\kern-.5em /}

\def \sgn {\mbox{\small sgn}}

\newcommand{\ci}[1]{}
\newcommand{\ba}{\begin{eqnarray}}
\newcommand{\ea}{\end{eqnarray}}
\newcommand{\bal}{\begin{align}}
\newcommand{\eal}{\end{align}}
\newcommand{\bay}[1]{\left(\begin{array}{#1}}
\newcommand{\eay}{\end{array}\right)}

\newcommand{\hide}[1]{}

\newcommand{\fsl}[1]{\ensuremath{\mathrlap{\!\not{\phantom{#1}}}#1}}

\newlist{axioms}{enumerate}{2}
\setlist[axioms,1]{label=\textbf{A\arabic{axiomsi}.}, ref=A\arabic{axiomsi}}
\setlist[axioms,2]{label=\textbf{A\arabic{axiomsi}\rlap{\myEnumCounter{axiomsii}}.},%
                   ref=A\arabic{axiomsi}\myEnumCounter{axiomsii},%
                   align=parleft,%
                   leftmargin=0em,%
                   itemsep=1.4ex,%
                   before={\stepcounter{axiomsi}}}

  \usetikzlibrary{decorations.markings}

\begin{document}

\begin{titlepage}
\begin{center}

\textbf{\LARGE
Lattice Chiral Fermion without Hermiticity
\vskip.3cm
}
\vskip .5in
{\large
Chen-Te Ma$^{a}$ \footnote{e-mail address: yefgst@gmail.com} 
and Hui Zhang$^{b,c}$ \footnote{e-mail address: Mr.zhanghui@m.scnu.edu.cn}
\\
\vskip 1mm
}
{\sl
%$^a$ 
%Department of Physics, Great Bay University, Dongguan, Guangdong 52300, China. 
%\\
$^a$ 
Department of Physics and Astronomy, Iowa State University, Ames, Iowa 50011, US. 
\\
$^b$
State Key Laboratory of Nuclear Physics and Technology,\\ 
Institute of Quantum Matter, South China Normal University, Guangzhou 510006, Guangdong, China.
\\
$^c$
Physics Department and Center for Exploration of Energy and Matter,\\
Indiana University, Bloomington, Indiana 47408, US.
}\\
\vskip 1mm
\vspace{40pt}
\end{center}

%\newpage
\begin{abstract} 
Our review of the lattice chiral fermion delves into some critical areas of lattice field theory.
By abandoning Hermiticity, the non-Hermitian formulation circumvents the Nielsen-Ninomiya theorem while maintaining chiral symmetry, a novel approach.
Comparing the Wilson and overlap fermions gives insight into how lattice formulations handle chiral symmetry.
The Wilson fermion explicitly breaks chiral symmetry to eliminate doublers. 
In contrast, the overlap fermion restores a modified form of chiral symmetry using the Ginsparg-Wilson relation.
We investigate how the (1+1)D Wilson fermion relates to the (1+1)D overlap fermion in the Hamiltonian formulation.
This connection could provide a clearer physical understanding of how chiral symmetry manifests at the lattice level.
Depending on Hermiticity for efficiency, Monte Carlo methods face unique challenges in a non-Hermitian setting.
We investigate how to correctly apply this method to non-Hermitian lattice fermions, which is essential for practical simulations.
Finally, the review of topological charge is crucial, as topological features in lattice formulations are strongly connected to chiral symmetry, anomalies, and the index theorem.
\end{abstract}
\end{titlepage}

\section{Introduction}
\label{sec:1}
\noindent
Lattice field theory indeed provides a robust framework for {\it non-perturbative} solutions to quantum field theory (QFT) \cite{Wilson:1974sk}.
The method, introduced by Wilson, replaces continuous spacetime with a discrete lattice, regularizing QFT \cite{Schwinger:1951ex} by introducing a natural cutoff—the lattice spacing—avoiding infinities that occur in the continuum path integral formulation \cite{Wilson:1974sk}.
In the continuum limit, where the lattice spacing tends to zero, and the lattice size becomes infinite, the theory should ideally converge to the desired continuum QFT \cite{Wilson:1974sk}.
\\

\noindent
One of the critical aspects of lattice field theory is that it makes renormalization possible in a non-perturbative setting by implementing the renormalization group (RG) \cite{Gell-Mann:1954yli} on a discrete space.
However, the connection between lattice computations and the continuum limit involves subtle issues.
The RG flow on the lattice may make it challenging to determine whether a lattice theory recovers the correct physics of the continuum theory.
For example, certain symmetries, such as chiral symmetry, are difficult to realize on the lattice without introducing either fermion doublers (as per the Nielsen-Ninomiya theorem) \cite{Nielsen:1980rz,Nielsen:1981xu} or using non-local operators (such as in overlap fermions) \cite{Neuberger:1998wv}.
Furthermore, the presence of lattice artifacts due to the finite lattice spacing requires careful extrapolation to the continuum limit, which can introduce computational challenges.
While lattice field theory regularizes QFT and simplifies numerical simulations, ensuring that the results correspond to physical continuum observables requires careful consideration of RG flows and lattice artifacts.
The problem is exceptionally intricate for theories with massless fermions or when maintaining chiral symmetry is crucial, as demonstrated by the developments around Wilson \cite{Wilson:1974sk} and overlap fermions.
\\

\noindent
Nielsen–Ninomiya theorem imposes significant constraints on lattice fermion formulations by showing that under certain conditions—namely, Hermiticity, locality, and chiral symmetry—it is impossible to avoid the fermion doubling problem when transitioning from a lattice to the continuum theory \cite{Karsten:1981gd}.
This implies that, under these assumptions, lattice fermions will always include multiple unwanted species (doublers), which is problematic for simulating models like the Standard Model.
To overcome this, different approaches relax one or more of these conditions:
\begin{itemize}
\item{Wilson fermions solves the doubling problem by explicitly breaking chiral symmetry.
A term is added to the lattice action that suppresses the doublers.
However, it introduces chiral symmetry breaking at {\it finite} lattice spacing, which needs to be carefully restored in the continuum limit.}
\item{Overlap fermions are based on the Ginsparg-Wilson relation \cite{Ginsparg:1981bj}, which modifies the concept of chiral symmetry on the lattice \cite{Luscher:1998pqa}.
Overlap fermions preserve a modified version of chiral symmetry \cite{Luscher:1998pqa} and can avoid doublers. However, they involve nonlocal operators \cite{Hernandez:1998et}, making them more computationally intensive.
}
\item{Tangent fermions avoid the doubling problem by breaking locality \cite{Beenakker:2023zel}.
Although the Hamiltonian features a highly non-local coupling between distant sites, it is possible to obtain the ground state from a local generalized eigenvalue problem \cite{Haegeman:2024qgf,Zakharov:2024xcg}. 
A similar transformation exists that illustrates the correspondence between a non-local Hermitian system and a local non-Hermitian system. This analogous concept can also be found in conformal field theories \cite{Guruswamy:1996rk,Hsin:2016blu,Hsieh:2022hgi}.
}
\end{itemize}
In our exploration of non-Hermitian formulations \cite{Stamatescu:1993ga}, it is interesting to consider how the relaxation of Hermiticity might allow different ways of dealing with fermion doubling and chiral symmetry without needing to follow the traditional paths of Wilson or overlap fermions.
\\

\noindent
We highlight an exciting connection between Wilson and overlap fermions, especially in (1+1)D Hamiltonian formulation \cite{Kogut:1974ag,Creutz:2001wp}.
The fact that on a finite lattice, the Wilson fermion formulation becomes equivalent to the overlap fermion formulation \cite{Hayata:2023zuk} suggests that certain lattice artifacts, typically problematic in lattice field theory, might manifest modified symmetries that reflect the underlying continuum symmetries.
This equivalence implies that the modified chiral symmetry of overlap fermions, which was introduced to bypass the no-go theorem (Nielsen-Ninomiya), also appears in the Wilson fermion formalism in such constrained scenarios.  
This discovery is insightful when exploring the broader symmetry structure of lattice field theories.
Since symmetries often dictate much of the physical content in lattice and continuum formulations, realizing a typical symmetry between two formulations is significant.
It hints that new lattice symmetries could be at play, possibly beyond the familiar continuum analogs, offering a pathway to further understanding how continuum physics can emerge cleanly from the lattice.
In particular, modified chiral symmetry on the lattice via the Ginsparg-Wilson relation in the overlap fermion case shows how lattice symmetries can evolve to include continuum-like behavior without directly imposing it.
Suppose this modified symmetry can be extended or generalized under certain conditions in the Wilson fermion formalism. 
In that case, it could lead to the discovery of more exotic symmetries inherent to the lattice regularization itself.
The (1+1)D case is a natural starting point for this exploration. 
However, extending these ideas to higher dimensions would be fascinating, potentially opening new avenues for both analytic and numerical investigation.
\\

\noindent
We dive into the subtle interplay between lattice regularization, chiral symmetry, and topological aspects, mainly through the lens of the Ginsparg-Wilson relation and its implications for {\it topological charge} and {\it zero modes}.
Finite lattices introduce artifacts that can break the symmetries of the underlying continuum theory.
Preserving these symmetries on the lattice improves the correspondence with the continuum, which is essential for analyzing numerical simulations.
However, a finite lattice generally depends on the lattice spacing, making preserving topological properties challenging.
The Atiyah-Singer index theorem connects the zero modes of the Dirac operator $D$ with the topological charge of the background gauge field \cite{Atiyah:1968mp,Atiyah:1970ws,Atiyah:1971rm}.
On the lattice, this theorem is non-trivial due to the discretization process \cite{Hasenfratz:1998ri}.
Nevertheless, efforts have been made to formulate the index theorem for lattice systems, especially in light of Ginsparg-Wilson fermions, which maintain a modified chiral symmetry.
By integrating out the fermion fields, one obtains a Jacobian that relates to the anomaly structure \cite{Fujikawa:1979ay} and the lattice index theorem \cite{Fujikawa:1998if,Adams:1998eg,Suzuki:1998yz}.
This approach helps formulate a lattice version of the index theorem.
Topological zero modes in the Dirac operator are a hallmark of topologically non-trivial gauge field backgrounds in the continuum.
However, on the lattice, the Ginsparg-Wilson relation does not always guarantee the presence of topological zero modes \cite{Chiu:1998bh} or a non-trivial topological charge \cite{Chiu:2001bg}.
This is due to the loss of the winding part of the topological charge density in the transition from the lattice to the continuum, even if the smooth part is recovered \cite{Chiu:2001ja}.
While the other part of the topological charge density can be recovered from topologically non-trivial gauge fields, the non-trivial contribution to the topological charge comes from the winding part \cite{Chiu:2001ja}.
This means that a lattice Dirac operator does not exhibit topological zero modes in such backgrounds \cite{Chiu:2001ja}.
While topological zero modes lack experimental evidence, the topological charge density has observable consequences, making it a more direct physical observable \cite{Chiu:2001ja}.
The subtlety arises from the fact that the topological charge density and zero modes are related in theory. 
The infinite size limit in the topological charge is subtle because because $\mathrm{Tr}(\gamma_5)=0$ loses the smooth continuum limit when the doublers disappear \cite{Fujikawa:1999ku}. 
However, the lattice formulation introduces complexities that decouple them in some cases.
In essence, we highlight how lattice formulations, particularly via the Ginsparg-Wilson relation, strive to capture continuum topological features. However, lattice artifacts and the system's finite nature create challenges in preserving the full topological structure, particularly in terms of zero modes and charge densities.
This remains an area of ongoing study and refinement.
\\

\noindent
The use of {\it one-sided} lattice differences in the non-Hermitian formulation {\it breaks} Hermiticity, which complicates conventional approaches to lattice fermion formulations \cite{Stamatescu:1993ga}.
This {\it breaking} of hypercubic symmetry means that the theory loses renormalizability in the standard sense, and physical observables should become dependent on the lattice spacing \cite{Stamatescu:1993ga}.
Non-physical poles in the propagator {\it decouple} in the continuum limit, leading to consistent particle descriptions as the spacing vanishes \cite{Stamatescu:1993ga}.
The loss of hypercubic symmetry at the lattice level can be addressed by performing a {\it quenched averaging} over all directions in the space \cite{Stamatescu:1993ga}.
This restores the symmetry at the level of the observables, as noted in 4D quantum electrodynamics (QED) studies, such as those in the weak-coupling expansion \cite{Sadooghi:1996ip}.
One key issue is that Monte Carlo simulations rely on a positive definite partition function, {\it not} necessarily on Hermiticity in the Lagrangian \cite{Guo:2021sjp}.
Introducing pseudo-fermion fields and considering fermions with one-sided differences in opposite directions can rewrite the theory to yield a positive definite partition function, making Monte Carlo (MC) simulations applicable \cite{Guo:2021sjp}.
This method has been demonstrated for free Dirac fermions \cite{Guo:2021sjp} and in models like the 2D Gross-Neveu-Yukawa (GNY) model \cite{Guo:2024jqt}.
Such an implementation is similar to investigating the Quantum Chromodynamics (QCD) phase diagram with opposite isospin chemical potential \cite{Brandt:2017oyy}.
The non-Hermitian formulation maintains chiral symmetry exactly, which leads to exciting properties regarding the index theorem.
The {\it bi-orthogonal} basis is introduced to realize the index theorem in a non-Hermitian setting \cite{Chen:2020kix}.
However, the {\it absence} of non-trivial topological charge, as seen in Hermitian formulations due to the exact chiral symmetry \cite{Bessho:2020hrs}, is a significant consequence \cite{Guo:2021sjp,Bessho:2020hrs}. 
One also realized the chiral fermion requiring the space-time reversal symmetry but sacrificing the Hermiticity \cite{Bender:2007nj,Chernodub:2017lmx,Ashida:2020dkc,Okuma:2022bnb}.
The symmetry of space-time reversal is spontaneously broken \cite{Chernodub:2017lmx}.
Therefore, preserving this symmetry does not help show the real spectrum when avoiding the Nielsen-Ninomiya theorem \cite{Chernodub:2017lmx}.
Despite the complexities introduced by the lack of Hermiticity, the non-Hermitian approach offers computational advantages.
For example, since the method {\it avoids} non-local operators, it can reduce simulation time.
Furthermore, preserving chiral symmetry at the lattice level provides convenience in analyzing the numerical result. 
\\

\noindent
The motivation for using lattice field theory is to study interacting field theories, particularly in the context of ultraviolet (UV) completeness and non-perturbative effects.
If a theory is not UV complete, the RG flow's consistency leads to triviality, implying the coupling vanishes \cite{Aizenman:1981du,Luscher:1987ay}.
Therefore, the practical application of lattice theory requires consideration of a UV complete theory.
Lattice QCD is a prominent example, where the asymptotic freedom of quarks and gluons at high energy scales \cite{Gross:1973id,Politzer:1973fx} justifies the use of lattice techniques.
The 2D GNY model also provides an excellent playground for investigating asymptotic safety, which was confirmed by the bosonization techniques (one flavor) \cite{Thirring:1958in,Skyrme:1958vn,Skyrme:1961vr,Coleman:1974bu,Mandelstam:1975hb,Buscher:1987sk,Buscher:1987qj,Burgess:1993np,Kovacs:2014fwa} and the lattice simulation and the resummation (two flavors) \cite{Guo:2024jqt}.
\\

\noindent
Lattice field theory is influential because it allows for non-perturbative calculations that are difficult to achieve using other techniques.
Resummation methods \cite{Ma:2022atx,Ma:2023uar} and lattice simulations all play roles in understanding field theory beyond perturbation theory.
However, the continuum limit remains a central challenge in lattice formulations, as correctly describing the original QFT without lattice artifacts is necessary.
\\

\noindent
The comparison between lattice results and non-perturbative resummation techniques highlights the importance of carefully examining non-perturbative phenomena.
In some cases, lattice methods can provide insights into invisible effects in perturbative expansions or requiring extensive resummation.
Balancing these methods to cross-check and validate results is essential, especially when studying models like the 2D GNY model where asymptotic safety or other non-perturbative behaviors are of interest.

\subsection{Outline}
\noindent
The outline of this review is as follows.
We introduce the path integral approach and its application in QFT in Sec.~\ref{sec:2} and the RG flow in Sec.~\ref{sec:3}.
The essential concepts of QFT related to the lattice field theory are covered.
\\

\noindent
We discuss the difficulties of putting the chiral fermions on a lattice and the various lattice constructions in Sec.~\ref{sec:4}.
It includes the challenges of defining chiral fermions on the lattice, including the Nielsen–Ninomiya theorem and the fermion doubling problem.
We review solutions like Wilson fermions, overlap fermions, and non-Hermitian formulations that circumvent the no-go theorem.
\\

\noindent
We present numerical methods for simulating lattice fermions in Sec.~\ref{sec:5}.
It includes discussions on the Metropolis and Hybrid MC algorithms and the conjugate gradient method, which is the most time-consuming part of simulating lattice fermions.
We discuss the chiral anomaly and the index theorem in Sec.~\ref{sec:6}.
It includes details about realizing the index theorem on the lattice, focusing on the Ginsparg-Wilson relation.
This section also includes connecting topological zero modes and non-trivial gauge backgrounds.
We outline potential research areas that could advance the study of lattice chiral fermions in Sec.~\ref{sec:7}.

\section{Path Integral Formulation}
\label{sec:2}
\noindent
We review the path integral formulation from the operator formulation.
We then provide an explicit example with a velocity-dependent potential to demonstrate the path integral formalism.
The subtlety of a measure is demonstrated in the finite temperature case, which is introduced by a Wick rotation.
We use the 0D QFT to show the necessity of the non-perturbative effect that the weak-coupling perturbation cannot reach.
Ultimately, we introduce the path integral formulation for the fermion fields.

\subsection{From Operator Formalism to Path Integral Formalism}
\noindent
We first derive the path integration formalism in 1D QFT.
In the path integral formalism, we use the Heisenberg picture.
The Heisenberg picture state $|x, t\rangle_H$ is an eigenstate of the Heisenberg picture position operator $X_H(t)$:
\bea
\hat{X}_H(t)|x, t\rangle_H\equiv x|x, t\rangle_H; \qquad \hat{X}_H(t)=e^{i \hat{H}t} \hat{X}_Se^{-i\hat{H}t}, 
\nn\\
\eea
where $\hat{X}_S$ is the time independent position operator, and $\hat{H}$ in the exponent is Hamiltonian.
The Schrödinger picture state is related to the Heisenberg picture state in the following way:
\bea
|x\rangle_S=e^{-i\hat{H}t}|x, t\rangle_H; \qquad |x\rangle_H=e^{i\hat{H}t}|x, t\rangle_S,
\eea
which is an eigenstate of the Schrödinger picture position operator
\bea
\hat{X}_S|x\rangle_S=x|x\rangle_S.
\eea
Because we want to show how operators become a classical variable in a path integral formalism, we use $\hat{}$ to label an operator for distinguishing operators and classical variables.
\\

\noindent
In classical mechanics, the momenta and position operators do not have an ambiguity for ordering, but lifting classical variables to quantum operators does.
The issue is due to the commutation relation
\bea
\lbrack \hat{p}, \hat{x}\rbrack=-i.
\eea
A classical Hamiltonian is
\bea
H(p, q)=f_{ab}p^aq^b,
\eea
and then a canonical quantization gives:
\bea
x\rightarrow \hat{x},\ p\rightarrow\hat{p}
\eea
with the commutation relation.
The quantum Hamiltonian is not unique like the following:
\bea
\hat{H}_L\equiv f_{ab}\hat{p}^a\hat{x}^b; \qquad \hat{H}_R\equiv f_{ab}\hat{x}^b\hat{p}^a.
\eea
We select either a Weyl ordering or a symmetric ordering
\bea
(\hat{x}\hat{p})_W\equiv\frac{1}{2}(\hat{p}\hat{x}+\hat{x}\hat{p})
\eea
for obtaining a hermitian Hamiltonian.
The transition amplitude from $x_I(t_I)$ to $x_F(t_F)$ is given by
\bea
{}_H\langle x_F, t_F|x_I, t_I\rangle_H={}_S\langle x_F|\hat{U}(t_F-t_I)|x_I\rangle_S,
\eea
where
\bea
\hat{U}(t)\equiv e^{-i\hat{H}t}
\eea
is a time-evolution operator.
We now divide a time interval as follows
\bea
t_n\equiv t_I+n\cdot\epsilon,
\eea
where $n=0, 1, \cdots, N$.
Therefore, we have a total of $N + 1$ points within the time interval, subject to the given conditions:
\bea
t_0\equiv t_I;\ t_N\equiv t_F;\ t_F-t_I=N\cdot\epsilon.
\eea
We then find that the amplitude becomes
\bea
&&{}_H\langle x_F, t_F|x_I, t_I\rangle_H
\nn\\
&=&
\lim_{N\rightarrow\infty}
\int dx_1dx_2\cdots dx_{N-1}\
\nn\\
&&
\times
{}_S\langle x_N|\hat{U}(\epsilon)|x_{N-1}\rangle_S{}_S\langle x_{N-1}|\hat{U}(\epsilon)|x_{N-2}\rangle_S\cdots{}_S\langle x_1|\hat{U}(\epsilon)|x_0\rangle_S.
\nn\\
\eea
We now calculate the transition amplitude over a short time interval:
\bea
&&
{}_S\langle x_{n+1}|\hat{U}_L(\epsilon)|x_n\rangle_S
\nn\\
&=&
{}_S\langle x_{n+1}\big|1-i\cdot\epsilon\hat{H_L}(\hat{p}_n, \hat{x}_n)\big|x_n\rangle_S+{\cal O}(\epsilon^2)
\nn\\
&=&\int dp_n\ {}_S\langle x_{n+1}|p_n\rangle_S{}_S\langle p_n\big|1-i\cdot\epsilon\hat{H_L}(\hat{p}_n, \hat{x}_n)\big|x_n\rangle_S
\nn\\
&&
+{\cal O}(\epsilon^2)
\nn\\
&=&
\frac{1}{\sqrt{2\pi}}\int dp_n\ e^{ip_nx_{n+1}}{}_S\langle p_n\big|1-i\cdot\epsilon\hat{H_L}(\hat{p}_n, \hat{x}_n)\big|x_n\rangle_S
\nn\\
&&
+{\cal O}(\epsilon^2)
\nn\\
&=&
\frac{1}{2\pi}\int dp_n\ e^{ip_n(x_{n+1}-x_n)}\big(1-i\cdot\epsilon H_L(p_n, x_n)\big)
+{\cal O}(\epsilon^2)
\nn\\
&=&
\int\frac{dp_n}{2\pi}\ \exp\bigg\lbrack i\epsilon\bigg(p_n\frac{x_{n+1}-x_n}{\epsilon}-H_L(p_n, x_{n})\bigg)\bigg\rbrack
\nn\\
&&
+{\cal O}(\epsilon^2),
\eea
where
\bea
{}_S\langle p_j|x_k\rangle{}_S\equiv\frac{1}{\sqrt{2\pi}}e^{-ip_jx_k}.
\eea
Here, we add a subscript to the time-evolution operator to denote that the Hamiltonian is $H_L$.
We can also derive the case for $\hat{U}_R(\epsilon)$ similarly.
The result is
\bea
&&
{}_S\langle x_{n+1}|\hat{U}_R(\epsilon)|x_n\rangle_S
\nn\\
&=&
\int\frac{dp_n}{2\pi}\ \exp\bigg\lbrack i\epsilon\bigg(p_n\frac{x_{n+1}-x_n}{\epsilon}-H_R(p_n, x_{n+1})\bigg)\bigg\rbrack
\nn\\
&&
+{\cal O}(\epsilon^2).
\eea
When we meet the momentum operators, we insert a complete basis of momenta.
Thus, the Hamiltonian is treated as a classical variable.
This is the fundamental concept behind the derivation.
Consequently, the Weyl ordering yields this result
\bea
&&
{}_S\langle x_{n+1}|\hat{U}_W(\epsilon)|x_n\rangle_S
\nn\\
&=&
\int\frac{dp_n}{2\pi}\ \exp\Bigg\lbrack i\epsilon\bigg\lbrack p_n\frac{x_{n+1}-x_n}{\epsilon}-H_W\bigg(p_n, \frac{x_{n+1}+x_n}{2}\bigg)\bigg\rbrack\Bigg\rbrack
\nn\\
&&
+{\cal O}(\epsilon^2).
\eea
We drop the subscript in the time-evolution operator because we will always use Weyl ordering later
\bea
&&{}_H\langle x_F, t_F|x_I, t_I\rangle_H
\nn\\
&=&\lim_{N\rightarrow\infty}
\int dx_1dx_2\cdots dx_{N-1}\frac{dp_0}{2\pi}\frac{dp_1}{2\pi}\cdots\frac{dp_{N-1}}{2\pi}\
\nn\\
&&\times
\exp\Bigg\lbrack i\sum_{n=0}^{N-1}\epsilon\bigg\lbrack p_n\frac{x_{n+1}-x_n}{\epsilon}-H\bigg(p_n, \frac{x_{n+1}+x_n}{2}\bigg)\bigg\rbrack\Bigg\rbrack
\nn\\
&&
+{\cal O}(\epsilon^2).
\eea
We can formally write the amplitude as
\bea
&&
{}_H\langle x_F, t_F|x_I, t_I\rangle_H
\nn\\
&&\equiv
\int\lbrack dx dp\rbrack\ \exp\bigg(i\int_{t_I}^{t_F}dt\ \big(p\dot{x}-H(p, x)\big)\bigg).
\eea
The path integral formalism is just a phase space integral.
The path integral formalism is useful because we only need to treat classical quantities using many properties of ordinary integrals.
It is convenient for the goal of calculating.
\\

\noindent
It is clear from various orderings that different discretizations yield different results.
Making different results makes sense because it encodes the information of the commutation relation between $\hat{p}$ and $\hat{x}$.
However, this makes the path integral more challenging than an ordinary integral
\bea
\lim_{N\rightarrow\infty}\sum_{n=0}^{N-1}\epsilon f(\xi_n)=\int_{t_I}^{t_F}dt\ f(t),\ \xi_n\in\lbrack t_n, t_{n+1}\rbrack.
\eea
In an ordinary integral, any value of $\xi_n$ within the interval $[t_n, t_{n+1}]$ can be used to obtain a result.
However, this is different for the path integral.
The primary reason for this difference is that the gap between $x_{n+1}$ and $x_n$ can be substantial even over a short time interval, as we consider all possible paths.
An evaluation of the path integral is necessary on a finite lattice with a discretization scheme.
The expression in a continuum limit is just formal for the convenience of writing due to the divergence.
We consistently refer back to a discretized form whenever we perform a calculation.
\\

\noindent
Now, let us examine the simplest Hamiltonian
\bea
H(p, x)=\frac{p^2}{2}+V(x).
\eea
The transition amplitude becomes
\bea
&&{}_H\langle x_F, t_F|x_I, t_I\rangle_H
\nn\\
&=&\lim_{N\rightarrow\infty}
\int dx_1dx_2\cdots dx_{N-1}\frac{dp_0}{2\pi}\frac{dp_1}{2\pi}\cdots\frac{dp_{N-1}}{2\pi}\
\nn\\
&&\times
\exp\Bigg\lbrack i\sum_{n=0}^{N-1}\epsilon\bigg\lbrack p_n\frac{x_{n+1}-x_n}{\epsilon}-\frac{p_n^2}{2}-V\bigg(p_n, \frac{x_{n+1}+x_n}{2}\bigg)\bigg\rbrack\Bigg\rbrack
\nn\\
&&
+{\cal O}(\epsilon^2).
\eea
We can find that the integration of momenta is just a Gaussian integral:
\bea
\int_{-\infty}^{\infty}dx\ e^{-ax^2+bx+c}=\sqrt{\frac{\pi}{a}}e^{\frac{b^2}{4a}+c};\ \mathrm{Re}(a)\ge0,\ a\neq 0.
\nn\\
\eea
Therefore, it is easy to evaluate.
We conclude that:
\bea
&&
\int\frac{dp_n}{2\pi}\ \exp\bigg(-\frac{i\epsilon}{2}p_n^2+i(x_{n+1}-x_n)p_n\bigg)
\nn\\
&&=
\frac{1}{\sqrt{2\pi i\epsilon}}\exp\bigg\lbrack\frac{i\epsilon}{2}\bigg(\frac{x_{n+1}-x_x}{\epsilon}\bigg)^2\bigg\rbrack;
\nn
\eea
\bea
&&
{}_H\langle x_F, t_F|x_I, t_I\rangle_H
\nn\\
&=&\lim_{N\rightarrow\infty}(2\pi i\epsilon)^{-\frac{N}{2}}\int dx_1dx_2\cdots dx_{N-1}
\nn\\
&&\times
\exp\Bigg\lbrack i\sum_{n=0}^{N-1}\epsilon\bigg\lbrack\frac{1}{2}\bigg(\frac{x_{n+1}-x_n}{\epsilon}\bigg)^2-V\bigg(\frac{x_{n+1}+x_n}{2}\bigg)\bigg\rbrack\Bigg\rbrack
\nn\\
&=&
\bigg(2\pi i\frac{t_F-t_I}{N}\bigg)^{-\frac{N}{2}}\int [dx]\ e^{i\int^{t_F}_{t_I}dt\ L(x, \dot{x})},
\eea
in which we use:
\bea
&&
\epsilon=\frac{t_F-t_I}{N}; \qquad \sum_{n=0}^{\infty}\epsilon\rightarrow\int_{t_I}^{t_F} dt;
\nn\\
&&
 \lim_{\epsilon\rightarrow 0}\frac{x_{n+1}-x_n}{\epsilon}\rightarrow\dot{x}.
\eea
Now, we obtain an integral path in a position space after integrating our momenta.
Because the normalization factor does not depend on the positions, the factor can be ignored or canceled for normalization when one calculates correlation functions.
However, the simple situation is not general after integrating out momenta.
We will give an exact case to show such a situation when the potential depends on a velocity.
However, the additional factor in the measure represents the quantum contribution to the action.

\subsection{Non-Linear Theory}
\noindent
Now we consider the Lagrangian
\bea
\label{nl}
L(x, \dot{x})=\frac{1}{2}\dot{x}^2f(x)-V(x),
\eea
which describes a system with a velocity-dependent potential.
The function $f(x)$ is smooth.
The momentum is:
\bea
p=\frac{\partial L}{\partial \dot{x}}=\dot{x}f(x).
\eea
Therefore, we derive the Hamiltonian:
\bea
H(p, x)&=&p\dot{x}-L=\frac{p^2}{f(x)}-\frac{p^2}{2f(x)}+V(x)
\nn\\
&=&\frac{p^2}{2f(x)}+V(x).
\eea
The transition amplitude is:
\bea
&&{}_H\langle x_F, t_F|x_I, t_I\rangle_H
\nn\\
&=&\lim_{N\rightarrow\infty}
\int dx_1dx_2\cdots dx_{N-1}\frac{dp_0}{2\pi}\frac{dp_1}{2\pi}\cdots\frac{dp_{N-1}}{2\pi}\
\nn\\
&&\times
\exp\Bigg\lbrack i\sum_{n=0}^{N-1}\epsilon\bigg\lbrack p_n\frac{x_{n+1}-x_n}{\epsilon}-\frac{p_n^2}{2f\big(\frac{x_{n+1}+x_n}{2}\big)}
\nn\\
&&
-V\bigg(\frac{x_{n+1}+x_n}{2}\bigg)\bigg\rbrack\Bigg\rbrack
+{\cal O}(\epsilon^2)
\nn\\
&=&\lim_{N\rightarrow\infty}(2\pi i\epsilon)^{-\frac{N}{2}}\int dx_1dx_2\cdots dx_{N-1}
\nn\\
&&\times
\exp\Bigg\lbrack i\sum_{n=0}^{N-1}\epsilon\bigg\lbrack\frac{1}{2}\bigg(\frac{x_{n+1}-x_n}{\epsilon}\bigg)^2f\bigg(\frac{x_{n+1}+x_n}{2}\bigg)
\nn\\
&&
-V\bigg(\frac{x_{n+1}+x_n}{2}\bigg)\bigg\rbrack\Bigg\rbrack
\nn\\
&&\times
\prod_{m=0}^{N-1}\sqrt{f\bigg(\frac{x_{m+1}+x_m}{2}\bigg)}
\nn\\
&=&\lim_{N\rightarrow\infty}(2\pi i\epsilon)^{-\frac{N}{2}}\int dx_1dx_2\cdots dx_{N-1}
\nn\\
&&\times
\exp\Bigg\lbrack i\sum_{n=0}^{N-1}\epsilon\bigg\lbrack\frac{1}{2}\bigg(\frac{x_{n+1}-x_n}{\epsilon}\bigg)^2f\bigg(\frac{x_{n+1}+x_n}{2}\bigg)
\nn\\
&&
-V\bigg(\frac{x_{n+1}+x_n}{2}\bigg)
-\frac{i}{2\epsilon}\ln f\bigg(\frac{x_{n+1}+x_n}{2}\bigg)\bigg\rbrack\Bigg\rbrack
\nn\\
&=&
\bigg(2\pi i\frac{t_F-t_I}{N}\bigg)^{-\frac{N}{2}}
\nn\\
&&\times
\int [dx]\ \exp\bigg\lbrack i\int^{t_F}_{t_I}dt\ \bigg(L(x, \dot{x})-\frac{i}{2\delta(0)}\ln f(x)\bigg)\bigg\rbrack,
\nn\\
\eea
in which we use:
\bea
&&
\epsilon=\frac{t_F-t_I}{N}; \qquad \sum_{n=0}^{\infty}\epsilon\rightarrow\int_{t_I}^{t_F} dt; 
\nn\\
&&
\frac{1}{\epsilon}\delta_{jk}\rightarrow\delta(t_j-t_k); \qquad \lim_{\epsilon\rightarrow 0}\frac{x_{n+1}-x_n}{\epsilon}\rightarrow\dot{x}
\eea
and one identity
\bea
\det A=e^{\mathrm{Tr}\ln A}.
\eea
Thus, by integrating out momenta, we obtain the Lagrangian and gain an additional term from the measure.

\subsection{Finite Temperature}
\noindent
The partition function at a finite temperature $T$ is
\bea
Z_{T}=\mathrm{Tr}\bigg(e^{-\beta H}\bigg),
\eea
where
\bea
\beta\equiv\frac{1}{T}
\eea
is the inverse temperature.
We can identify $\beta$ as $it$.
In a 1D bosonic quantum field theory, the path integral at finite temperature is defined as follows
\bea
Z_{T}=\int\lbrack d\phi\rbrack\ \exp\big(-S_T(\phi)\big),
\eea
where
\bea
S_T(\phi)\equiv\int_0^{\beta}d\tau\ L_E(\phi).
\eea
The $L_E$ is the Lagrangian for the Euclidean time.
The scalar field satisfies the periodic boundary condition
\bea
\phi(\tau)=\phi(\tau+\beta).
\eea
If the Lagrangian for the scalar field theory is
\bea
L=-\frac{1}{2}\partial_t\phi\partial^t\phi-V(\phi),
\eea
the action at a finite temperature becomes
\bea
S_T(\phi)=\int_0^{\beta}d\tau\ \bigg(\frac{1}{2}(\partial_{\tau}\phi)^2+V(\phi)\bigg).
\eea
In QFT at finite temperature, the process is equivalent to performing a Wick rotation, defined as $t\equiv -i\tau$.
The path integral at finite temperature is similar to doing a path integral with finite time.

\subsection{Measure}
\noindent
Now, we examine the issue of a measure in scalar field theory with a potential
\bea
V(\phi)=\frac{1}{2}m^2\phi^2
\eea
at finite temperature.
The scalar field satisfies the equation of motion
\bea
(-\partial_{\tau}^2+m^2)\phi(\tau)=\lambda\cdot\phi(\tau).
\eea
We write $\phi(\tau)$ as
\bea
\phi(\tau)=\sum_n\phi_n\cdot f_n(\tau),
\eea
Because $\phi(\tau)$ is a real quantum field, the $\phi_n$ has the constraint
\bea
\phi_n^*=\phi_{-n}.
\eea
where $n$ is an integer.
Then the $f_n(\tau)$ can be
\bea
f_n(\tau)=\frac{1}{\sqrt{\beta}}\exp\bigg(\frac{2\pi i n\tau}{\beta}\bigg).
\eea
Therefore, we can determine the eigenvalue of $f_n(\tau)$
\bea
\lambda_n=m^2+\frac{4\pi^2n^2}{\beta^2}.
\eea
In other words, we solve the equation
\bea
(-\partial_{\tau}^2+m^2)f_n(\tau)=\lambda_n\cdot f_n(\tau),
\eea
and the relation between the $\lambda$ and $\lambda_n$ is
\bea
\lambda=\sum_n\lambda_n.
\eea
The $f_n(\tau)$ satisfies the orthogonal condition
\bea
\int_0^{\beta}d\tau\ f_n(\tau)^*f_m(\tau)=\delta_{nm}.
\eea
The orthogonal condition implies
\bea
\phi_n=\int_0^{\beta}d\tau\ f_n^*(\tau)\phi(\tau).
\eea
\\

\noindent
The action, written in terms of modes, is
\bea
S_E=\frac{1}{2}\sum_n\lambda_n\phi^*_n\phi_n.
\eea
To normalize the partition function
\bea
Z_T=\int\lbrack d\phi\rbrack\ \exp(-S_E)=1,
\eea
the measure $\lbrack d\phi\rbrack$ is defined as
\bea
\lbrack d\phi\rbrack\equiv\prod_n\bigg(\sqrt{\frac{\lambda_n}{\pi}}\cdot d\phi_n\bigg).
\eea
Hence we obtain
\bea
\langle e^{-r\cdot S_E}\rangle=\prod_n\bigg(\frac{1}{\sqrt{1+r}}\bigg)=0,
\eea
where $r$ is an arbitrary positive number.
It is easy to know that the configuration with an infinite action vanishes.
Therefore, this implies that the configuration with a finite action also vanishes.
We only have two possibilities for such a situation.
The first situation is no finite action.
This situation is generally impossible because soliton solutions exist in QFT.
Therefore, we anticipate a scenario in which a finite action has no measurable impact.
The zero measure does not contribute to a path integral.
Now, let us explore the partition function.
The partition function is normalized, and all integration ranges extend from $-\infty$ to $\infty$.
We first define
\bea
\sqrt{\lambda_n}\phi_n\equiv a_n.
\eea
We can write the expression of the partition function as the following:
\bea
Z_T=\prod_n\lim_{L\rightarrow\infty}\int_{-L}^{L}\frac{da_n}{\sqrt{\pi}}\ e^{-\frac{1}{2}a_n^2}=1.
\eea
If we change the ordering of the infinite limitation and infinite products, we obtain
\bea
\lim_{L\rightarrow\infty}\prod_n\int^L_{-L}\frac{da_n}{\sqrt{\pi}}\ e^{-\frac{1}{2}a_n^2}=\prod_n A_n(L)=0,
\eea
where $A_n(L)<1$.
Before we take $L\rightarrow\infty$, the infinite products of $A_n(L)$ already vanish.
Hence, we obtain the same conclusion.
The configuration contributes to path integration only when the action becomes infinite (as $L\rightarrow\infty$).
In other words, the finite action has zero measure.
\\

\noindent
Let us use the discretized version to revisit the issue
\bea
\frac{1}{2}\int dt\ \dot{\phi}^2\rightarrow \frac{1}{2}\sum_j \frac{(\phi_{j+1}-\phi_j)^2}{\epsilon}.
\eea
The infinite action at least requires
\bea
\bigg\langle\frac{(\phi_{j+1}-\phi_j)^2}{\epsilon}\bigg\rangle\sim 1.
\eea
This results in a configuration that is not smooth
\bea
\dot{\phi}^2\sim\frac{1}{\epsilon}.
\eea
The kinematic energy approaches infinity under the limit $\epsilon\rightarrow 0$.
A configuration contributes to path integration only when it is not differentiable.

\subsection{Perturbation Issue}
\noindent
The partition function of 0D QFT is just an integration:
\bea
Z(g)=\int^{\infty}_{-\infty}\frac{d\phi}{\sqrt{2\pi}}\ e^{-\frac{\phi^2}{2}-g\frac{\phi^4}{24}}
=\sqrt{\frac{3}{2\pi g}}e^{\frac{3}{4g}}K_{\frac{1}{4}}\bigg(\frac{3}{4g}\bigg),
\nn\\
\eea
where $K_{\nu}(x)$ is the modified Bessel function of the second kind
\bea
K_{\nu}(z)\equiv\frac{\Gamma\big(\nu+\frac{1}{2}\big)(2z)^{\nu}}{\sqrt{\pi}}\int_0^{\infty}dt\ \frac{\cos t}{(t^2+z^2)^{\nu+\frac{1}{2}}}.
\eea
We can also use the weak-coupling perturbation to calculate the integration:
\bea
\frac{1}{\sqrt{2\pi}}\int_{-\infty}^{\infty}d\phi\ e^{-\frac{\phi^2}{2}}&=&1;
\nn\\
-g\frac{1}{4!\sqrt{2\pi}}\int_{-\infty}^{\infty}d\phi\ e^{-\frac{\phi^2}{2}}\phi^4&=&-\frac{g}{8};
\nn\\
(-g)^2\frac{1}{2!(4!)^2\sqrt{2\pi}}\int_{-\infty}^{\infty}d\phi\ e^{-\frac{\phi^2}{2}}\phi^8&=&\frac{35g^2}{384};
\nn\\
(-g)^3\frac{1}{3!(4!)^3\sqrt{2\pi}}\int_{-\infty}^{\infty}d\phi\ e^{-\frac{\phi^2}{2}}\phi^{12}&=&-\frac{385g^3}{3072};
\nn\\
(-g)^4\frac{1}{4!(4!)^4\sqrt{2\pi}}\int_{-\infty}^{\infty}d\phi\ e^{-\frac{\phi^2}{2}}\phi^{16}&=&\frac{25025g^4}{98304};
\nn\\
(-g)^5\frac{1}{5!(4!)^5\sqrt{2\pi}}\int_{-\infty}^{\infty}d\phi\ e^{-\frac{\phi^2}{2}}\phi^{20}&=&-\frac{1616615g^5}{2359296};
\nn\\
(-g)^6\frac{1}{6!(4!)^6\sqrt{2\pi}}\int_{-\infty}^{\infty}d\phi\ e^{-\frac{\phi^2}{2}}\phi^{24}&=&\frac{260275015g^6}{113246208};
\nn\\
(-g)^7\frac{1}{7!(4!)^7\sqrt{2\pi}}\int_{-\infty}^{\infty}d\phi\ e^{-\frac{\phi^2}{2}}\phi^{28}&=&-\frac{929553625g^7}{100663296};
\nn
\eea
\bea
&&
\ln Z(g)
\nn\\
&=&-\frac{g}{8}+\frac{g^2}{12}-\frac{11g^3}{96}+\frac{17g^4}{72}
\nn\\
&&
-\frac{619g^5}{960}+\frac{709g^6}{324}-\frac{858437g^7}{96768}+{\cal O}(g^8). 
\eea
Thus, the coefficient increases rapidly as the powers of $g$ rise.
Indeed, the perturbation series of the $\ln Z(g)$ is not convergent, and the coefficient of the $g^m$ grows as $m!$.
The series is only asymptotically convergent, meaning that including the first few terms at a small value of $g$ produces an exact solution with good precision.
Developing non-perturbative techniques to study a quantum system is essential because the weak-coupling perturbation expansion can be problematic not only in strong coupling regions but also in weak coupling regions due to the issue of non-convergence.

\subsection{Fermion Fields}
\noindent
The fermion field is a Grassmann variable in a path formalism.
The Grassmann variables satisfy the following algebra
\bea
\theta_j\theta_k=-\theta_k\theta_j.
\eea
Therefore, the algebra shows:
\bea
\theta_j^2=-\theta_j^2=0
\eea
for each $j$.
This indicates that Grassmann variables cannot be considered numbers.
Consider the below $c$-function
\bea
f(\eta)=a+\alpha\eta,
\eea
where $a$ is a $c$-number, $\alpha$ is a Grassmann constant, and $\eta$ is a Grassmann variable.
\\

\noindent
Now, we will discuss the integration of Grassmann variables.
We first assume that the integration has some elementary properties:
\bea
\int d\eta\ \big(b_1f(\eta)+b_2g(\eta)\big)&=&b_1\int d\eta\ f(\eta)+b_2\int d\eta\ g(\eta); 
\nn\\
\int d\eta\ \frac{\partial}{\partial\eta}f(\eta)&=&0,
\eea
where $b_1$ and $b_2$ are $c$-constants.
Requiring that the integration is translational invariance
\bea
\int d\eta\ f(\eta)=\int d\eta\ f(\eta+\beta),
\eea
where $\beta$ is a Grassmann constant, we obtain:
\bea
a\int d\eta +\alpha\int d\eta\ \eta&=&\int d\eta\ (a+\alpha\eta+\alpha\beta)
\nn\\
&=&(a+\alpha\beta)\int d\eta +\alpha\int d\eta\ \eta.
\eea
As a result, we derive additional rules for integration.:
\bea
\int d\eta=0; \qquad
\int d\eta\ \eta=1.
\eea
\\

\noindent
Now we consider the Grassmann variables and their complex conjugation, and we obtain
\bea
\int d\eta^* d\eta\ \exp(-\eta^*A\eta)=\det (A).
\eea
We show the calculation steps for the constant value for $A$:
\bea
\int d\eta^*d\eta\ \exp(-\eta^*A\eta)&=&\int d\eta^*d\eta\ (1-\eta^*A\eta)
\nn\\
&=&\int d\eta^*d\eta\ (1+\eta\eta^*A)
\nn\\
&=&A.
\eea
We find that integrating bosonic variables differs significantly from integrating Grassmann variables.

\section{RG Flow}
\label{sec:3}
\noindent
We introduce the concepts of RG flow \cite{Gell-Mann:1954yli} beginning from quantum mechanics (QM).
We then discuss the relations between the physical, renormalized, and bare parameters in 4D $\lambda\phi^4$ theory.
We present the Wilson and Polchinski exact renormalization group equation (ERGE) and discuss the triviality in 4D $\lambda\phi^4$ theory \cite{Aizenman:1981du} from the Polchinski ERGE.

\subsection{QM}
\noindent
The initial concept of RG flow indicates that parameters are influenced by regularization.
The regularization is to modify a theory at a scale of a cut-off to a well-defined level.
After the modification, the theory has been made more consistent and well-defined.
We demonstrate the concept from QM
\bea
-\frac{1}{2}\nabla^2\Psi(x)+c\cdot\delta^2(x)\Psi(x)=E\Psi(x),
\eea
where $E$ is the eigenenergy, the dimensionless parameter $c$ does not depend on the spatial coordinates, and
we solve the above equation from the spherically symmetric case
Hence, the Hamiltonian becomes
\bea
-\frac{1}{2r}\frac{d}{dr}\bigg(r\frac{d\Psi}{dr}\bigg)+\frac{c}{2\pi r}\delta(r-a)\Psi(r)=E\Psi(r),
\eea
in which we used:
\bea
\nabla^2\Psi&=&\frac{1}{r}\frac{d}{dr}\bigg(r\frac{d\Psi}{dr}\bigg); 
\nn\\
\delta^2(x)&\equiv&\frac{1}{2\pi r}\delta(r)\rightarrow \frac{1}{2\pi r}\delta(r-a).
\eea
When $r\neq a$, the wavefunction is a linear combination of the Bessel function of the first kind $J_0(pr)$ and the Bessel function of the second kind $Y_0(pr)$.
The Bessel functions for the integer $n$ are defined as:
\bea
J_{n}(x)&\equiv&\frac{1}{\pi}\int_0^{\pi}d\tau\ \cos\big(n\tau-x\cdot\sin(\tau)\big); 
\nn\\
Y_n(x)&\equiv&\lim_{\alpha\rightarrow n}\frac{J_{\alpha}(x)\cos(\alpha\pi)-J_{-\alpha}(x)}{\sin(\alpha\pi)}.
\eea
Therefore, the wavefunction is expressed as
\bea
\Psi(r)=\left\{\begin{array}{ll}
A\cdot J_0(pr)+B\cdot Y_0(pr) & r>a, \\
C\cdot J_0(pr), & r<a,
\end{array} \right.
\eea
where
\bea
p=\sqrt{2E}.
\eea
Then, we multiply $r$ by the equation
\bea
-\frac{1}{2}\frac{d}{dr}\bigg(r\frac{d\Psi}{dr}\bigg)+\frac{c}{2\pi}\delta(r-a)\Psi(r)=E\cdot r\Psi(r).
\eea
Then we integrate the equation for the $r$ from $a-\epsilon$ to $a+\epsilon$ and take the limit $\epsilon\rightarrow 0$ in the final:
\bea
&&\lim_{\epsilon\rightarrow 0}\int_{a-\epsilon}^{a+\epsilon}dr\ \bigg\lbrack-\frac{1}{2}\frac{d}{dr}\bigg(r\frac{d\Psi}{dr}\bigg)+\frac{c}{2\pi}\delta(r-a)\Psi(r)\bigg\rbrack
\nn\\
&=&-\frac{a}{2}\big(\Psi^{\prime}(a+\epsilon)-\Psi^{\prime}(a-\epsilon)\big)+\frac{c}{2\pi}\Psi(a);
\nn
\eea
\bea
E\lim_{\epsilon\rightarrow 0}\int_{a-\epsilon}^{a+\epsilon}dr\ r\Psi(r)=0;
\eea
\bea
\Psi^{\prime}(a+\epsilon)-\Psi^{\prime}(a-\epsilon)=\frac{c}{\pi a}\Psi(a).
\eea
In this context, we assume that the wavefunction is continuous at $r = a$, but its derivative is not.
\\

\noindent
We now have two conditions: a continuous wavefunction and a discontinuous derivative to solve the equation of QM.
When $r\gg a$, the wavefunction's asymptotic behavior is:
\bea
\Psi(r)
&&
\rightarrow A\cdot \cos\bigg(pr-\frac{\pi}{4}\bigg)+B\cdot\sin\bigg(pr-\frac{\pi}{4}\bigg)
\nn\\
&&
\propto\cos\bigg(pr-\frac{\pi}{4}+\delta_0\bigg),
\eea
where
\bea
\tan\delta_0\equiv-\frac{A}{B}.
\eea
The coefficients $A$ and $B$ can be rewritten in terms of $C$:
\bea
A&=&\Bigg\lbrack 1-\frac{c}{\pi}\bigg\lbrack\ln\bigg(\frac{pa}{2}\bigg)+\gamma\bigg\rbrack+{\cal O}(p^2a^2)\Bigg\rbrack C;
\nn\\
B&=&\bigg(\frac{c}{2}+{\cal O}(p^2a^2)\bigg)C,
\eea
where $\gamma$ is the Euler-Mascheroni constant
\bea
\gamma=\lim_{n\rightarrow\infty}\bigg(-\ln n+\sum_{k=1}^n\frac{1}{k}\bigg),
\eea
for
\bea
p\ll\frac{1}{a}.
\eea
Hence we obtain
\bea
\cot\delta_0=-\frac{2}{c}+\frac{2}{\pi}\bigg\lbrack\ln\bigg(\frac{pa}{2}\bigg)+\gamma\bigg\rbrack
+{\cal O}(p^2a^2).
\eea
It is easy to observe that the right-hand side depends on the regularization parameter $a$.
However, the $\delta_0$ is a physical observable.
Therefore, the value of $c$ must also depend on the value of $a$.
\\

\noindent
We now identify the regularization parameter as the momentum cutoff
\bea
a\equiv\frac{1}{\Lambda}.
\eea
Then we obtain:
\bea
0&=&\frac{d}{d\Lambda}\cot\delta_0
=\bigg(\frac{\partial}{\partial\Lambda}+\frac{dc}{d\Lambda}\frac{\partial}{\partial c}\bigg)\cot\delta_0
\nn\\
&=&\frac{2}{\pi}\frac{2\Lambda}{p}\bigg(-\frac{p}{2\Lambda^2}\bigg)+\frac{dc}{d\Lambda}\frac{2}{c^2},
\eea
which leads to
\bea
\Lambda\frac{d}{d\Lambda}\bigg(\frac{1}{c}\bigg)=-\frac{1}{\pi}.
\eea
Therefore, the parameter is essential for the regularization process.
The theoretical consistency requires that the regularization influence the parameter.
Consequently, the RG flow is also necessary in QM.
The parameter $c$ is called the bare parameter.
We also use an energy scale $\mu$ to define the renormalized parameter
\bea
\frac{1}{c_R(\mu)}\equiv\frac{1}{c(\Lambda)}-\frac{1}{\pi}\bigg\lbrack\ln\bigg(\frac{\mu}{2\Lambda}\bigg)+\gamma\bigg\rbrack.
\eea
A physical observable should not depend on the energy scale $\mu$.
Now we can rewrite the equation in terms of the renormalized parameter
\bea
\cot\delta_0=-\frac{2}{c_R(\mu)}+\frac{2}{\pi}\ln\frac{p}{\mu}+{\cal O}\bigg(\frac{p^2}{\Lambda^2}\bigg).
\eea
Requiring that the $\delta_0$ does not depend on the $\mu$, and then it shows
\bea
\mu\frac{\partial}{\partial\mu}\bigg(\frac{1}{c_R}\bigg)=-\frac{1}{\pi}.
\eea
We are more interested in the renormalized parameter because it is closely related to the physical observable
\bea
\cot\delta_0=-\frac{2}{c_R(\mu=p)}+{\cal O}\bigg(\frac{p^2}{\Lambda^2}\bigg).
\eea
The equation indicates that the momentum cut-off must exceed the energy scale.
Otherwise, the running picture will become invalid.
When the energy scale is reduced, maintaining the same low-energy physics requires adjusting the coupling constant.

\subsection{4D $\lambda\phi^4$ Theory}
\noindent
We consider the 4D $\lambda\phi^4$ theory as an example.
The Lagrangian is
\bea
L=-\frac{1}{2}\partial_{\mu}\phi\partial^{\mu}\phi-\frac{1}{2}(m^2-i\epsilon)\phi^2
-\frac{\lambda}{4!}\phi^4.
\eea
Now, let us proceed to calculate the two-point function.
\bea
G_2(x, y)\equiv\langle\phi(x)\phi(y)\rangle.
\eea
We do an expansion up to the first-order in $\lambda$,
\bea
&&
G_2(x, y)
\nn\\
&=&\frac{1}{Z[\lambda]}\int{\cal D}\phi\ e^{iS\lbrack\lambda=0\rbrack}\phi(x)\phi(y)
\nn\\
&&\times
\bigg\lbrack 1+\int d^4\tilde{x}\ \bigg(-i\frac{\lambda}{4!}\phi^4(\tilde{x})\bigg)\bigg\rbrack+\cdots,
\eea
where $Z[\lambda]$ is the partition function, and $S\lbrack\lambda\rbrack$ is the action.
When one also expands the partition function in $\lambda$ at the leading order, one can apply Wick's theorem to the calculation of correlation functions:
\bea
&&
\frac{1}{Z\lbrack\lambda=0\rbrack}\int{\cal D}\phi\ e^{iS\lbrack\lambda=0\rbrack}\phi(x)\phi(y)
\nn\\
&&\times
\bigg\lbrack 1+\int d^4\tilde{x}\ \bigg(-i\frac{\lambda}{4!}\phi^4(\tilde{x})\bigg)\bigg\rbrack
\nn\\
&=&
\frac{1}{Z\lbrack\lambda=0\rbrack}\int{\cal D}\phi\ e^{iS\lbrack\lambda=0\rbrack}\phi(x)\phi(y)
\nn\\
&&
+
\frac{1}{Z\lbrack\lambda=0\rbrack}\int{\cal D}\phi\ e^{iS\lbrack\lambda=0\rbrack}\phi(x)\phi(y)\int d^4\tilde{x}\ \bigg(-i\frac{\lambda}{4!}\phi^4(\tilde{x})\bigg)
\nn\\
&=&
\tilde{G}_2(x, y)
-\frac{i\lambda}{2}\int d^4\tilde{x}\ \tilde{G}_2(\tilde{x}, \tilde{x})
\tilde{G}_2(x, \tilde{x})\tilde{G}_2(y, \tilde{x})
\nn\\
&&
+\cdots,
\eea
where
\bea
\tilde{G}_2(x, y)\equiv i\int\frac{d^4p}{(2\pi)^4}\frac{e^{ip(x-y)}}{-p^2-m^2+i\epsilon}.
\eea
\\

\noindent
The $\cdots$ will be eliminated through the expansion of the partition function
\bea
-\frac{\lambda}{\big\lbrack Z\lbrack \lambda=0\rbrack\big\rbrack^2}\delta Z\int{\cal D}\phi\ e^{iS\lbrack\lambda=0\rbrack}\phi(x)\phi(y),
\eea
in which $\delta Z$ is defined by:
\bea
Z\lbrack\lambda\rbrack&\equiv& Z\lbrack\lambda=0\rbrack+\lambda\cdot \delta Z+\cdots; 
\nn\\
\delta Z&=&\int{\cal D}\phi\ e^{iS\lbrack\lambda=0\rbrack}\int d^4\tilde{x} \bigg(\frac{-i}{4!}\phi^4(\tilde{x})\bigg).
\eea
The integration in the momentum space is given by:
\bea
&&-\frac{i\lambda}{2}\int d^4\tilde{x}\ \tilde{G}_2(\tilde{x}, \tilde{x})
\tilde{G}_2(x, \tilde{x})\tilde{G}_2(y, \tilde{x})
\nn\\
&=&
-\frac{i\lambda}{2}\int d^4\tilde{x}\frac{d^4p_1}{(2\pi)^4}\frac{d^4p_2}{(2\pi)^4}\frac{d^4p_3}{(2\pi)^4}\
\nn\\
&&\times
\frac{i}{-p_1^2-m^2+i\epsilon}\frac{i}{-p_2^2-m^2+i\epsilon}\frac{i}{-p_3^2-m^2+i\epsilon}
\nn\\
&&\times
e^{i(p_2x+p_3y)}e^{-i\tilde{x}(p_2+p_3)}
\nn\\
&=&
-\frac{i\lambda}{2}\int \frac{d^4p_1}{(2\pi)^4}\frac{d^4p_2}{(2\pi)^4}\frac{d^4p_3}{(2\pi)^4}\
\nn\\
&&\times
\frac{i}{-p_1^2-m^2+i\epsilon}\frac{i}{-p_2^2-m^2+i\epsilon}\frac{i}{-p_3^2-m^2+i\epsilon}
\nn\\
&&\times
e^{i(p_2x+p_3y)}
(2\pi)^4\delta^4(p_2+p_3)
\nn\\
&=&
-\frac{i\lambda}{2}\int \frac{d^4p_1}{(2\pi)^4}\frac{d^4p_2}{(2\pi)^4}\
\nn\\
&&\times
\frac{i}{-p_1^2-m^2+i\epsilon}\frac{i}{-p_2^2-m^2+i\epsilon}\frac{i}{-p_3^2-m^2+i\epsilon}
\nn\\
&&\times 
e^{ip_2(x-y)}.
\eea
Consequently, the two-point function in momentum space takes the following form
\bea
&&
\bar{G}_2(p)
\nn\\
&=&\frac{i}{-p^2-m^2+i\epsilon}
\nn\\
&&
-\frac{i\lambda}{2}\bigg(\frac{i}{-p^2-m^2+i\epsilon}\bigg)^2\int\frac{d^4p_1}{(2\pi)^4}\frac{i}{-p_1^2-m^2+i\epsilon}
\nn\\
&&
+\cdots.
\eea
The quantum correction of the mass term is given by:
\bea
&&
\frac{i}{-p^2-m^2-\delta m^2+i\epsilon}
\nn\\
&=&
\frac{i}{(-p^2-m^2+i\epsilon)\big(1-\frac{\delta m^2}{-p^2-m^2+i\epsilon}\big)}
\nn\\
&=&
\frac{i}{-p^2-m^2+i\epsilon}+\frac{i\delta m^2}{(-p^2-m^2+i\epsilon)^2}
\nn\\
&&
+\cdots.
\eea
Hence the $\delta m^2$ is
\bea
\delta m^2=i\frac{\lambda}{2}\int\frac{d^4p}{(2\pi)^4}\frac{1}{-p^2-m^2+i\epsilon}.
\eea
\\

\noindent
Now we do a Wick rotation for the momentum
\bea
p^0=ip_{E}^0.
\eea
The Euclidean metric is $\mathrm{diag}(+, +, \cdots, +)$.
Then we obtain:
\bea
\delta m^2&=&i\frac{\lambda}{2}\int\frac{d^4p}{(2\pi)^4}\frac{1}{-p^2-m^2+i\epsilon}
\nn\\
&=&-\frac{\lambda}{2}\int\frac{d^4p_{E}}{(2\pi)^4}\frac{1}{-p_{E}^2-m^2+i\epsilon}
\nn\\
&=&\frac{\lambda}{2}\int\frac{d^4p_{E}}{(2\pi)^4}\frac{1}{p_{E}^2+m^2-i\epsilon}.
\eea
We redefine the momentum as the below:
\bea
p_{1, E}&\equiv& p_E\sin(\theta_1)\sin(\theta_2)\cos(\phi);
\nn\\
p_{2, E}&\equiv& p_E\sin(\theta_1)\sin(\theta_2)\sin(\phi);
\nn\\
p_{3, E}&\equiv& p_E\sin(\theta_1)\cos(\theta_2);
\nn\\
p_{4, E}&\equiv& p_E\cos(\theta_1).
\eea
The range of the variables is provided as follows:
\bea
0\le\theta_1\le\pi;\ 0\le\theta_2\le\pi;\ 0\le\theta_3\le2\pi;\ 0\le p_E\le\Lambda.
\nn\\
\eea
\\

\noindent
Calculating the Jacobian matrix obtains:
\bea
&&
\int d^4p_{1, E}
\nn\\
&=&\int_0^{\pi} d\theta_2\ \sin^2(\theta_2)\int_0^{\pi}d\theta_1\ \sin(\theta_1)
\int_0^{2\pi}d\phi\int_{0}^{\Lambda}dp_{E}\ p_E^3
\nn\\
&=&2\pi^2\int_{0}^{\Lambda}dp_{E}\ p_E^3.
\nn\\
\eea
As a result, we can conclude the following:
\bea
\delta m^2
&=&
\frac{\lambda}{2}\int\frac{d^4p_{1, E}}{(2\pi)^4}\frac{1}{p_{1, E}^2+m^2-i\epsilon}
\nn\\
&=&
\frac{\lambda}{16\pi^2}\int_{0}^{\Lambda}dp_E\ p_E^3\frac{1}{p_{E}^2+m^2-i\epsilon}
\nn\\
&=&
\frac{\lambda}{32\pi^2}\int_{0}^{\Lambda^2}du\ \frac{u}{u+m^2-i\epsilon}
\nn\\
&=&\frac{\lambda}{32\pi^2}\int_{0}^{\Lambda^2}du\ \bigg(1-\frac{m^2-i\epsilon}{u+m^2-i\epsilon}\bigg)
\nn\\
&=&
\frac{\lambda}{32\pi^2}\bigg\lbrack\Lambda^2-(m^2-i\epsilon)
\ln\bigg(1+\frac{\Lambda^2}{m^2-i\epsilon}\bigg)\bigg\rbrack,
\nn\\
\eea
where
\bea
u\equiv p_E^2.
\eea
Indeed, we use the property of contour integration and assume some conditions.
When the poles enclosed in the closed loops are the same, the value of the contour integration is not changed.
We assume that the contour integral approaches zero at the boundary.
We used the property and assumption to evaluate the above integration.
The integration becomes divergent when the momentum approaches infinity.
It is why we put a cut-off $\Lambda$ in the momentum.
The procedure is regularization.
We will now discuss the assumptions related to the vanishing boundary integration.
Indeed, the divergent boundary integration is not due to the time component of the momentum variable.
More precisely, the contour integration for the time component of the momentum variable without including the integration for the spatial components of the momentum variable has the vanishing boundary integration.
To begin with, we perform a Wick rotation for the time component of the momentum variable and then proceed to integrate the spatial components of the momentum variable.
Consequently, applying Wick's rotation only necessitates zero boundary integral of the time-component momentum variable.
\\

\noindent
Because the physical mass
\bea
m_{\mathrm{phys}}^2\equiv m^2+\delta m^2
\eea
is independent of the cut-off:
\bea
\frac{d}{d\Lambda}m_{\mathrm{phys}}^2=\frac{d}{d\Lambda}(m^2+\delta m^2)=0,
\eea
we obtain
\bea
\Lambda\frac{d m^2}{d\Lambda}&=&-\frac{\lambda}{16\pi^2}
\bigg(\Lambda^2-(m^2-i\epsilon)\frac{\Lambda^2}{m^2+\Lambda^2-i\epsilon}\bigg)
\nn\\
&&
+\cdots,
\eea
where $\cdots$ represents contributions from higher-order effects.
Now, we will select a significantly high cutoff to achieve
\bea
\Lambda\frac{d m^2}{d\Lambda}\approx-\frac{\lambda}{16\pi^2}
\bigg(\Lambda^2-(m^2-i\epsilon)\bigg)+\cdots.
\eea
We can then solve the flow equation
\bea
m^2(\Lambda)\approx-\frac{\lambda}{32\pi^2}\bigg(\Lambda^2-m^2\ln\frac{\Lambda^2}{\mu^2}\bigg)+m_r^2,
\eea
in which the variables, energy scale $\mu$ and renormalized mass $m_r$, depend on the initial conditions of the flow equation,
to obtain the bare mass:
\bea
m^2(\Lambda)&\approx&\frac{m_r^2-\frac{\lambda}{32\pi^2}\Lambda^2}{1-\frac{\lambda}{32\pi^2}\ln\frac{\Lambda^2}{\mu^2}}
\nn\\
&\approx&
m_r^2\bigg(1+\frac{\lambda}{32\pi^2}\ln\frac{\Lambda^2}{\mu^2}\bigg)
-\frac{\lambda}{32\pi^2}\Lambda^2.
\eea
Therefore, the physical mass can be expressed as:
\bea
m_{\mathrm{phys}}^2
=
m^2+\delta m^2
=
m_r^2\bigg(1+\frac{\lambda}{32\pi^2}\ln\frac{m_r^2}{\mu^2}\bigg).
\eea
The physical mass is not affected by the momentum cut-off.
When we choose $m_r^2=\mu^2$, the square of physical mass is $\mu^2$ again.
Thus, we can see that the renormalized mass closely relates to the physical observable, which is the physical mass.
The observation appeared in QM.
We now utilize QFT to revisit this observation once more.

\subsection{Wilson ERGE}
\noindent
The Wilson ERGE aims to create an adequate low-energy description by integrating a high-energy mode and maintaining the same partition function.
We begin from a partition function with a momentum cut-off $\Lambda$ and integrate out a high-energy mode:
\bea
Z&=&\int_{p\le\Lambda}{\cal D}\phi\ \exp\big(-S_{\Lambda}(\phi)\big)
\nn\\
&=&\int_{p\le\tilde{\Lambda}}{\cal D}\phi_-\int_{\tilde{\Lambda}\le p\le\Lambda}{\cal D}\phi_+\ \exp\big(-S_{\Lambda}(\phi)\big)
\nn\\
&=&\int_{p\le\tilde{\Lambda}}{\cal D}\phi_-\ \exp\big(-S_{\tilde{\Lambda}}(\phi)\big),
\eea
where
\bea
\exp\big(-S_{\tilde{\Lambda}}(\phi)\big)\equiv\int_{\tilde{\Lambda}\le p\le\Lambda}{\cal D}\phi_+\ \exp\big(-S_{\Lambda}(\phi)\big).
\nn\\
\eea
We can determine that the operation is transitive.
If we calculate $S_{\tilde{\Lambda}}$ from the $S_{\Lambda}$ and then calculate $S_{\bar{\Lambda}}$ from the $S_{\tilde{\Lambda}}$, it is equivalent to calculating $S_{\bar{\Lambda}}$ from the $S_{\Lambda}$.
Therefore, the transition indicates that the operation is associative
\bea
O_2\circ(O_1\circ S_{\Lambda})=(O_2\circ O_1)\circ S_{\Lambda}.
\eea
The $O_1\circ S_{\Lambda}$ means the calculation of $S_{\tilde{\Lambda}}$ from the $S_{\Lambda}$.
The $O_2\circ(O_1\circ S_{\Lambda})$ means the calculation of $S_{\bar{\Lambda}}$ from the $S_{\tilde{\Lambda}}$.
The $(O_2\circ O_1)\circ S_{\Lambda}$ means the calculation of $S_{\bar{\Lambda}}$ from the $S_{\Lambda}$.
Therefore, the Wilson ERGE is a semigroup.
Because the operation for integrating out a high-energy mode does not have an inverse operation, Wilson ERGE is not a group.
The Wilson ERGE offers a physical perspective for a theory with a cut-off, but its practical application is challenging.
We will later introduce the Polchinski ERGE to simplify the calculations.

\subsection{Polchinski ERGE}
\noindent
Now, we will illustrate how Polchinski discusses the ERGE in the context of scalar field theory.
The Euclidean Lagrangian is:
\bea
L_E&=&\frac{1}{2}\partial_{\mu}\phi\partial^{\mu}\phi+V(\phi)
\nn\\
&\equiv&\frac{1}{2}\partial_{\mu}\phi\partial^{\mu}\phi
+\sum_{k=1}^{\infty}\Lambda^{d-k(d-2)}\frac{\lambda_{2k}}{(2k!)}\phi^{2k},
\eea
where $d$ is the dimensions of spacetime.
In this discussion, we focus on dimensionality greater than two, as operators involving derivatives, aside from the kinematic term, become irrelevant.
The relevant operators characterize our macroscopic world.
The coupling constants $\lambda_{2k}$ are dimensionless.
We also require the symmetry $\phi\rightarrow-\phi$ in the Euclidean Lagrangian.
Therefore, we can utilize the Euclidean Lagrangian to analyze Polchinski's ERGE in scalar field theory.
We separate the scalar field from low- and high-energy modes as the following
\bea
\phi=\Phi+\chi.
\eea
Now we choose $\Phi$ as a constant background and then integrate out the high-energy mode $\chi$.
The quadratic effective action is given by
\bea
&&
S_E\lbrack\Phi+\chi\rbrack-S_E\lbrack\Phi\rbrack
\nn\\
&=&
\int d^dx\
\bigg(\frac{1}{2}\partial_{\mu}\chi\partial^{\mu}\chi+\frac{1}{2}V^{\prime\prime}(\Phi)\chi^2\bigg)+\cdots,
\eea
in which $\cdots$ is the higher-order term of $\chi$.
After the Fourier transformation, the quadratic action becomes:
\bea
&&
\int d^dx\
\bigg(\frac{1}{2}\partial_{\mu}\chi\partial^{\mu}\chi+\frac{1}{2}V^{\prime\prime}(\phi)\chi^2\bigg)
\nn\\
&=&
\int_{\Lambda-\delta\Lambda<|p|\le\Lambda}\frac{d^dp}{(2\pi)^d}\
\bar{\chi}(p)\bigg(\frac{1}{2}p^2+\frac{1}{2}V^{\prime\prime}(\Phi)\bigg)\bar{\chi}(-p)
\nn\\
&=&\frac{\Lambda^{d-1}\delta\Lambda}{2(2\pi)^d}\big(\Lambda^2+V^{\prime\prime}(\Phi)\big)
\int_{S^{d-1}}d\Omega\ \bar{\chi}(\Lambda\hat{p})\bar{\chi}(-\Lambda\hat{p}),
\nn\\
\eea
where $\hat{p}$ is a unit vector along the radial direction of a sphere.
We select a small variation of the momentum cut-off.
This facilitates the smooth integration of high-energy modes, simplifying the calculation of an RG flow.
This is the main technique used in the Polchinski ERGE.
\\

\noindent
Now, we will integrate the high-energy mode to determine the variation of the cut-off in the effective action
\bea
e^{-\delta_{\Lambda}S_{\mathrm{eff}}}\sim
\big(\Lambda^2+V^{\prime\prime}(\Phi)\big)^{-\frac{N}{2}},
\eea
where $N$ is the number of momentum modes in the narrow shell.
We do a regularization by placing the scalar field theory in a box with a size $L$ and periodic boundary conditions.
The momentum is quantized as follows
\bea
p_{\mu}\equiv\frac{2\pi}{L}n_{\mu},
\eea
where $n_{\nu}$ is an arbitrary integer.
Therefore, the number of momentum modes is given by
\bea
N=L^d\frac{\mathrm{Vol}(S^{d-1})}{(2\pi)^d}\Lambda^{d-1}\delta\Lambda.
\eea
Hence we obtain
\bea
\delta_{\Lambda}S_{\mathrm{eff}}=a\Lambda^{d-1}\delta\Lambda\int d^dx\ \ln\big(\Lambda^2+V^{\prime\prime}(\Phi)\big),
\eea
where
\bea
a\equiv\frac{\mathrm{Vol}(S^{d-1})}{2(2\pi)^d}=\frac{1}{(4\pi)^{\frac{d}{2}}\Gamma\big(\frac{d}{2}\big)},
\eea
and the integration comes from the factor $L^d$.
\\

\noindent
We combine the action and its correction and do the derivative for the momentum cut-off to obtain
\bea
\Lambda\frac{dS_{\mathrm{eff}}}{d\Lambda}+\Lambda\frac{\delta_{\Lambda}S_{\mathrm{eff}}}{\delta\Lambda}=0,
\eea
which leads to the following:
\bea
&&0
\nn\\
&=&
\sum_{k=1}^{\infty}\big(d-(k(d-2)\big)\Lambda^{d-k(d-2)}\frac{\lambda_{2k}}{(2k)!}\Phi^{2k}
\nn\\
&&
+\sum_{k=1}^{\infty}\Lambda^{d-k(d-2)}\frac{\Phi^{2k}}{(2k)!}\Lambda\frac{d\lambda_{2k}}{d\Lambda}
\nn\\
&&
+a\Lambda^d\sum_{k=1}^{\infty}\bigg(\frac{1}{(2k)!}\frac{d^{2k}}{d\Phi^{2k}}
\ln\big(\Lambda^2+V^{\prime\prime}(\phi)\big)\bigg)\bigg|_{\Phi=0}\Phi^{2k};
\nn
\eea
\bea
&&
\Lambda\frac{d\lambda_{2k}}{d\Lambda}
\nn\\
&=&\big(k(d-2)-d\big)\lambda_{2k}
\nn\\
&&
-a\Lambda^{k(d-2)}\frac{d^{2k}}{d\Phi^{2k}}\ln\big(\Lambda^2+V^{\prime\prime}(\Phi)\big)\bigg|_{\Phi}=0.
\eea
The terms for $k=1$ and $k=2$ give:
\bea
\Lambda\frac{d\lambda_2}{d\Lambda}&=&-2\lambda_2-a\frac{\lambda_4}{1+\lambda_2}; 
\nn\\
\Lambda\frac{d\lambda_4}{d\Lambda}&=&(d-4)\lambda_4
-a\frac{\lambda_6}{1+\lambda_2}
+3a\frac{\lambda_4^2}{(1+\lambda_2)^2}.
\eea
\\

\noindent
We now assume that the coupling constants are weak, allowing us to linearize the flow equations as follows:
\bea
\beta_{2k}=\Lambda\frac{d \lambda_{2k}}{d\Lambda}=\big(k(d-2)-d\big)\lambda_{2k}-a\lambda_{2k+2}.
\eea
Hence we obtain a solution
\bea
\lambda_{2k}=0
\eea
for $k>1$.
This solution presents the most straightforward form of the fixed-point, known as the Gaussian fixed-point.
We observe that the $\phi^2$ term is relevant for $k=1$ for a non-interacting theory.
As soon as \(\lambda_2\) becomes sufficiently large, the reliability of the approximation is compromised.
Indeed, the exact solution shows that the $\beta_{2k}$ approaches $\big(k(d-2)-d\big)\lambda_{2k}$ or the quantum correction is suppressed.
\\

\noindent
Now we consider $d=4$, the coupling constants flow as the below:
\bea
\Lambda\frac{d\lambda_2}{d\Lambda}&=&-2\lambda_2-a\frac{\lambda_4}{1+\lambda_2}; 
\nn\\
\Lambda\frac{d\lambda_4}{d\Lambda}&=&
-a\frac{\lambda_6}{1+\lambda_2}
+3a\frac{\lambda_4^2}{(1+\lambda_2)^2}.
\eea
For the coupling constants $\lambda_6$, $\lambda_8$, $\cdots$, etc., these represent irrelevant operators.
Therefore, we can disregard these operators.
The flow equations become:
\bea
\Lambda\frac{d\lambda_2}{d\Lambda}&=&-2\lambda_2-a\frac{\lambda_4}{1+\lambda_2}; 
\nn\\
\Lambda\frac{d\lambda_4}{d\Lambda}&\approx&
3a\frac{\lambda_4^2}{(1+\lambda_2)^2}
=\frac{3}{16\pi^2}\lambda_4^2.
\eea
Therefore, the solution for $\lambda_4$ is as follows
\bea
\frac{1}{\lambda_4}=-\frac{3}{16\pi^2}\ln\Lambda+C_{\Lambda},
\eea
where $C_{\Lambda}$ is an arbitrary constant.
We choose
\bea
C_{\Lambda}=\frac{3}{16\pi^2}\ln\mu
\eea
to obtain
\bea
\lambda_4(\Lambda)=\frac{16\pi^2}{3\ln\big(\frac{\mu}{\Lambda}\big)}.
\eea
The $\phi^4$ term is the highest power of $\phi$ in the relevant operators of 4D scalar field theory.
For a consistent theory, it requires $\lambda_4>0$.
Otherwise, the theory is not bounded when $|\phi|\rightarrow\infty$.
This limits the available options
\bea
\mu>\Lambda.
\eea
Once we have obtained $\beta_4$, we can determine the flow of the coupling constant.
We need to worry about the divergence for the $\beta_4$ because it occurs at a finite scale $\Lambda=\mu$.
This indicates that we do not have a non-trivial interacting QFT based on this result.
To rely on the perturbation results in the 4D $\lambda\phi^4$ theory, we must select the cut-off scale and coupling constant to be as small as possible.

\section{Lattice Formulations}
\label{sec:4}
\noindent
We discuss the fermion doubling issue arising from naive lattice fermions and the more general Nielsen–Ninomiya no-go theorem \cite{Nielsen:1980rz,Nielsen:1981xu,Karsten:1981gd}.
The most common approach is to abandon (Wilson fermion \cite{Wilson:1974sk}) or modify (overlap fermion \cite{Neuberger:1998wv}) the chiral symmetry.
We present these approaches from the Wilson and overlap fermions.
In particular, these two approaches are equivalent in the (1+1)D Hamiltonian formulation \cite{Kogut:1974ag,Creutz:2001wp,Hayata:2023zuk}.
We then present a non-Hermitian lattice formulation that bypasses the no-go theorem by relaxing the requirement of Hermiticity \cite{Stamatescu:1993ga}.

\subsection{Fermion Doubling}
\noindent
We introduce the doubling problem from the 1D Dirac fermion field.
The Euclidean action is
\bea
S_D&=&\int dx\ \bar{\psi}(x)(\gamma_1\partial_1+m)\psi(x)
\nn\\
&\equiv&
\int dx\ \bar{\psi}(x)(D+m)\psi(x)
,
\eea
where
\bea
\gamma_1\equiv\sigma_3=
\begin{pmatrix}
1&0
\\
0&-1
\end{pmatrix}; \
\bar{\psi}\equiv\psi^{\dagger}\gamma_1;
\partial_1\equiv\frac{\partial}{\partial x}
.
\eea
The $D$ is called the Dirac matrix.
The propagator satisfies
\bea
\bigg(\gamma_1\frac{d}{dx}+m\bigg)S(x)=\delta(x).
\eea
The solution is:
\bea
S(x)&=&\int_{-\infty}^{\infty}\frac{dp}{2\pi}\ e^{ipx}\frac{1}{i\gamma_1p+m}
\nn\\
&=&
\int_{-\infty}^{\infty}\frac{dp}{2\pi}\ e^{ipx}\frac{-i\gamma_1p+m}{p^2+m^2}
\nn\\
&=&\bigg(-\gamma_1\frac{d}{dx}+m\bigg)\int_{-\infty}^{\infty}\frac{dp}{2\pi}\ e^{ipx}\frac{1}{p^2+m^2}
\nn\\
&=&\bigg(-\gamma_1\frac{d}{dx}+m\bigg)\frac{1}{2m}e^{-m|x|}
\nn\\
&=&\big(\gamma_1\cdot\sgn(x)+1\big)\frac{1}{2}e^{-m|x|}
\nn\\
&=&
\begin{pmatrix}
\theta(x)&0
\\
0&\theta(-x)
\end{pmatrix}
e^{-m|x|},
\eea
where
\bea
\theta(x)\equiv\Bigg\{\begin{array}{ll}
1, & x\ge 0 \\
0, & x<0
\end{array}.
\eea
\\

\noindent
A naive discretization for the Euclidean action is:
\bea
S_{DL}&=&a\sum_{n=0}^{N-1}\bar{\psi}(n)\bigg(\gamma_1\frac{\psi(n+1)-\psi(n-1)}{2a}+m\psi(n)\bigg)
\nn\\
&=&a\sum_{n_1, n_2; \alpha_1, \alpha_2}\bar{\psi}(n_1)_{\alpha_1}\big(D(n_1, n_2)_{\alpha_1, \alpha_2}
\nn\\
&&
+m\delta_{\alpha_1, \alpha_2}\delta_{n_1, n_2}\big)
\psi(n_2)_{\alpha_2},
\eea
where
\bea
D(n_1, n_2)_{\alpha_1, \alpha_2;}\equiv
(\gamma_1)_{\alpha_1, \alpha_2}\frac{\delta_{n_1+1, n_2}-\delta_{n_1-1, n_2}}{2a}.
\nn\\
\eea
The matrix components of $\gamma_1$ are labeled by $\alpha_1, \alpha_2=1, 2$.
The number of lattice points is given by $N$.
The lattice fermion field satisfies the anti-periodic boundary condition
\bea
\psi(0)=-\psi(N).
\eea
Therefore, the propagator on the momentum space satisfies the following equations:
\bea
\bigg(\gamma_1\frac{i\sin(pa)}{a}+m\bigg)\bar{S}(p)=1; \qquad
\bar{S}(p)=\frac{1}{\gamma_1\frac{i\sin(pa)}{a}+m}.
\nn\\
\eea
When the number of lattice points goes to infinity, the propagator in the position space becomes:
\bea
S_n&=&\int^{\frac{\pi}{a}}_{-\frac{\pi}{a}}\frac{dp}{2\pi}\ e^{ipna}\frac{1}{\gamma_1\frac{i\sin(pa)}{a}+m}
\nn\\
&=&\int^{\frac{\pi}{a}}_{-\frac{\pi}{a}}\frac{dp}{2\pi}\ e^{ipna}
\begin{pmatrix}
\frac{a}{i\sin(pa)+ma}&0
\\
0&\frac{a}{-i\sin(pa)+ma}
\end{pmatrix}.
\nn\\
\eea
After integrating the momentum, the first diagonal element is
\bea
S_{11}(x)=\Bigg\{\begin{array}{ll}
\frac{\big(-ma+\sqrt{1+m^2a^2}\big)^n}{\sqrt{1+m^2a^2}}, & x=na>0 \\
(-1)^{|n|}\frac{\big(-ma+\sqrt{1+m^2a^2}\big)^{|n|}}{\sqrt{1+m^2a^2}}, & x=na<0
\end{array};
\nn\\
\eea
The second diagonal element is
\bea
S_{22}(x)=S_{11}(-x).
\eea
\\

\noindent
Now we take the continuum limit $ma\rightarrow0$ with a fixed $x=na$ and then obtain
\bea
S_{11}(x)\rightarrow\Bigg\{\begin{array}{ll}
e^{-m|x|}, & x>0 \\
(-1)^{|n|}e^{-m|x|}, & x<0
\end{array},
\eea
in which we used
\bea
\lim_{ma\rightarrow 0}\bigg(1-ma\bigg)^{\frac{1}{ma}}=e.
\eea
It is easy to find that the propagator does not return to the expected continuum theory for $x<0$.
The reason is that we have two poles, which are at:
\bea
\gamma_1\sin(pa)=ima; \qquad \gamma_1\sin(\pi-pa)=ima.
\eea
The appearance of the unwanted pole is due to the sine function.
The non-perturbative effect of the lattice spacing non-trivially introduces the non-physical degrees to the fermion fields.
In a continuum theory, we only have one pole in any dimension.
When considering $d$ dimensions, we have $2^d-1$ unwanted poles.
\\

\noindent
The Nielsen-Ninomiya no-go theorem demonstrates that chiral symmetry and Fermion doubling cannot be simultaneously avoided when placing a Dirac fermion field on a lattice.\cite{Nielsen:1980rz,Nielsen:1981xu,Karsten:1981gd}:
\begin{itemize}
\item{1. $D(x)$ is a local operator, which means it is bounded by $\sim\exp(-|x|/c)$, where $c$ is proportional to the lattice spacing $a$.;
}
\item{2. $\bar{D}(p)=i\gamma_{\mu}p_{\mu}+{\cal O}(ap^2)$ for $p\ll\pi/a$;
}
\item{3. $\bar{D}(p)$ is invertible for $p\neq 0$ (no massless doublers);
}
\item{4. $\gamma_5D+D\gamma_5=0$ (chiral symmetry),
}
\end{itemize}
where the spacetime indices are denoted by $\mu$.
The common approach is to abandon or modify chiral symmetry on a lattice.
The chiral transformation is
\bea
\psi\rightarrow e^{i\alpha\gamma_5}\psi; \qquad
\bar{\psi}\rightarrow\bar{\psi}e^{i\alpha\gamma_5},
\eea
where $\alpha$ is a constant.
The $\gamma_5$ satisfies the algebra
\bea
\gamma_{\mu}\gamma_5+\gamma_5\gamma_{\mu}=0.
\eea
A chiral invariant Lagrangian must satisfy the following condition:
\bea
L_F=\bar{\psi}D\psi=\bar{\psi}e^{i\alpha\gamma_5}De^{i\alpha\gamma_5}\psi.
\eea
This condition is equivalent to the condition
\bea
D\gamma_5+\gamma_5D=0.
\eea

\subsection{Wilson Fermion}
\noindent
To solve the doubling issue, we can introduce the Wilson term \cite{Wilson:1974sk}
\bea
&&
W_L
\nn\\
&=&-a\frac{\delta_{a_1, a_2}\delta_{n_1+1, n_2}-2\delta_{a_1, a_2}\delta_{n_1, n_2}+\delta_{a_1, a_2}\delta_{n_1-1, n_2}}{2a^2}
\nn\\
\eea
to generate an infinite mass for the unwanted pole at the continuum limit \cite{Wilson:1974sk}.
The Wilson term at a continuum limit is
\bea
W_L\sim-\frac{a}{2}\partial_1\partial_1.
\eea
Because the kinematic term only has one derivative, it is easy to know that a higher derivative term should vanish at a continuum limit.
The propagator is \cite{Wilson:1974sk}
\bea
S_{DL}=\int^{\frac{\pi}{a}}_{-\frac{\pi}{a}}\frac{dp}{2\pi}e^{ipna}\frac{1}{\gamma_1\frac{i\sin(pa)}{a}+m+\frac{1}{a}\big(1-\cos(pa)\big)}.
\nn\\
\eea
Therefore, we can determine that the mass of the unwanted pole is given by the expression $m+2/a$ \cite{Wilson:1974sk}.
This mass becomes infinitely heavy in the continuum limit.
When we take the continuum limit, the unwanted pole becomes irrelevant.
However, the massless fermion makes reaching an expected continuum theory hard, and the Wilson fermion does not preserve chiral symmetry in a finite lattice \cite{Wilson:1974sk}.

\subsection{Overlap Fermion}
\noindent
We first introduce the lattice chiral symmetry \cite{Luscher:1998pqa} given by the Ginsparg-Wilson relation \cite{Ginsparg:1981bj}.
We then discuss the spectrum of the Dirac operator satisfying the Ginsparg-Wilson relation.
Ultimately, we offer a solution, the overlap fermion.

\subsubsection{Lattice Chiral Symmetry}
\noindent
Indeed, the lattice artifact can modify the symmetry of a lattice.
Hence, modifying the chiral transformation on a finite lattice is fine.
We adopt the chiral transformation \cite{Luscher:1998pqa}:
\bea
\psi&\rightarrow&\exp\bigg\lbrack i\alpha\gamma_5\bigg(1-\frac{a}{2}D\bigg)\bigg\rbrack\psi; 
\nn\\
\bar{\psi}&\rightarrow&\bar{\psi}\exp\bigg\lbrack i\alpha\bigg(1-\frac{a}{2}D\bigg)\gamma_5\bigg\rbrack.
\eea
When the fermion field interacts with other fields, the chiral transformation on a lattice can depend on the fields.
The continuum transformation does not depend on fields due to the coupling.
Hence, the chiral symmetry of a lattice fermion field is different from that of the continuum case.
\\

\noindent
Due to the chiral transformation, the chiral invariant Lagrangian requires the  condition
\bea
&&
\bar{\psi}D\psi
\nn\\
&=&
\bar{\psi}\exp\bigg\lbrack i\alpha\bigg(1-\frac{a}{2}D\bigg)\gamma_5\bigg\rbrack D
\exp\bigg\lbrack i\alpha\gamma_5\bigg(1-\frac{a}{2}D\bigg)\bigg\rbrack\psi.
\nn\\
\eea
The above equality can be held:
\bea
&&\bar{\psi}\exp\bigg\lbrack i\alpha\bigg(1-\frac{a}{2}D\bigg)\gamma_5\bigg\rbrack D
\exp\bigg\lbrack i\alpha\gamma_5\bigg(1-\frac{a}{2}D\bigg)\bigg\rbrack\psi
\nn\\
&=&\bar{\psi}\exp\bigg\lbrack i\alpha\bigg(1-\frac{a}{2}D\bigg)\gamma_5\bigg\rbrack
\exp\bigg\lbrack -i\alpha\bigg(1-\frac{a}{2}D\bigg)\gamma_5\bigg\rbrack D\psi
\nn\\
&=&
\bar{\psi}D\psi
\eea
when $D$ satisfies
\bea
D\gamma_5\bigg(1-\frac{a}{2}D\bigg)+\bigg(1-\frac{a}{2}D\bigg)\gamma_5 D=0,
\eea
which is equivalent to the Ginsparg-Wilson relation \cite{Ginsparg:1981bj}
\bea
D\gamma_5+\gamma_5D=aD\gamma_5D.
\eea

\subsubsection{Spectrum of Dirac Operator}
\noindent
Now, we will discuss the eigenvalues of the Dirac operator
\bea
D v_{\lambda}=\lambda v_{\lambda},
\eea
when the operator satisfies the Ginsparg-Wilson relation.
The Dirac operator is $\gamma_5$-Hermitian
\bea
\gamma_5 D\gamma_5=D^{\dagger},
\eea
which leads to the following:
\bea
\det(D-\lambda)
&=&
\det\big(\gamma_5^2(D-\lambda)\big)
=
\det\big(\gamma_5(D-\lambda)\gamma_5)
\nn\\
&=&
\det(D^{\dagger}-\lambda)
=
\big(\det(D-\lambda^*)\big)^*.
\eea
The findings suggest that the eigenvalues are either real numbers or occur in pairs of complex conjugates:
\bea
\lambda(v_{\lambda}, \gamma_5 v_{\lambda})
&=&
(v_{\lambda}, \gamma_5 Dv_{\lambda})
=
(v_{\lambda}, D^{\dagger}\gamma_5v_{\lambda})
\nn\\
&=&
(D v_{\lambda}, \gamma_5v_{\lambda})
=\lambda^*(v_{\lambda}, \gamma_5v_{\lambda}),
\eea
in which the inner product of two vectors, $u_1$ and $u_2$, are defined by
\bea
(u_1, u_2)\equiv u_1^{\dagger}u_2.
\eea
Hence, this directly implies
\bea
(v_{\lambda}, \gamma_5v_{\lambda})=0
\eea
when $\lambda$ is not real-valued.
In other words, we have the non-vanishing chirality
\bea
(v_{\lambda}, \gamma_5 v_{\lambda})\neq 0
\eea
for real eigenvalues.
\\

\begin{figure}[!htb]
%\begin{center}
\includegraphics[width=1.\textwidth]{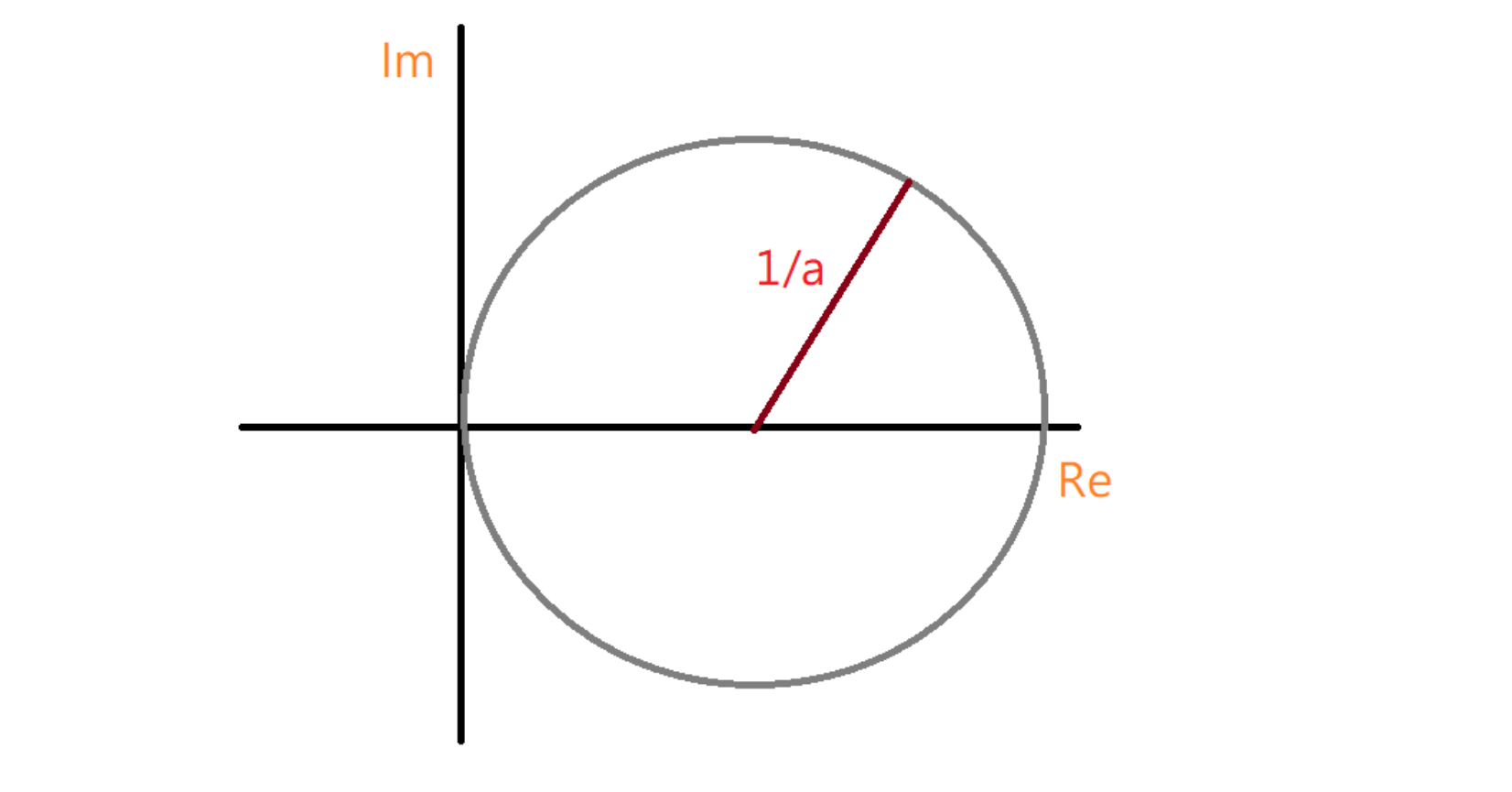}
\caption{
The real and imaginary parts of the eigenvalues lie on a circle centered at $x=1/a$, $y=0$, with a radius of $1/a$.
}
%\end{center}
\label{GWcircle}
\end{figure}

\noindent
The Ginsparg-Wilson relation implies that $D$ is a normal operator:
\bea
DD^{\dagger}=D^{\dagger}D=D+D^{\dagger}.
\eea
The eigenvectors form an orthogonal basis.
The eigenvalues follow the equation
\bea
\lambda^*+\lambda=a\lambda^*\lambda.
\eea
When the definition of an eigenvalue is as follows
\bea
\lambda=x+iy,
\eea
we can obtain the Ginsparg-Wilson circle (Fig. \ref{GWcircle})
\bea
\bigg(x-\frac{1}{a}\bigg)^2+y^2=\frac{1}{a^2}.
\eea
It is easy to know that we have two real eigenvalues:
\bea
0;\ \frac{2}{a}.
\eea
The unwanted mode, $2/a$, will decouple from the continuum limit.
\\

\noindent
When the eigenvector corresponds to a zero mode, we obtain the following:
\bea
&&
D v_0=0
\nn\\
&\rightarrow& 
0=\gamma_5Dv_0=(aD\gamma_5D-D\gamma_5)v_0=-D\gamma_5v_0.
\nn\\
\eea
Since the Dirac operator acting on the zero-mode eigenvector commutes with $\gamma_5$, one can select zero modes as the eigenstates of $\gamma_5$.
Since
\bea
\gamma_5^2=1,
\eea
we obtain
\bea
\gamma_5v_0=\pm v_0.
\eea
This indicates that the zero modes exhibit chirality.
The right-handed mode has the positive chirality, and the left-handed mode has the negative chirality.

\subsubsection{Overlap Dirac Operator}
\noindent
Now we show one solution, overlap Dirac operator \cite{Neuberger:1998wv} to the Ginsparg-Wilson relation:
\bea
D_{ov}&=&\frac{1}{a}\big(1+\gamma_5\mathrm{sgn}(Q)\big)=\frac{1}{a}\bigg(1+\gamma_5Q(Q^2)^{-\frac{1}{2}}\bigg);
\nn\\
 Q&=&\gamma_5A,
\eea
where $A$ is some suitable operator with the $\gamma_5$-Hermitian.
Because the operator $A$ satisfies the $\gamma_5$-Hermitian, this shows that $Q$ is Hermitian, which implies only real eigenvalues,
\bea
Q^{\dagger}=A^{\dagger}\gamma_5=\gamma_5A\gamma_5\gamma_5=\gamma_5A=Q.
\eea
One choice for $A$ is \cite{Neuberger:1998wv}
\bea
A=aD_W-(1+s),
\eea
where $D_W$ is the Wilson-Dirac operator, and $|s|$ is bounded by one
\bea
|s|<1.
\eea
The overlap Dirac operator becomes \cite{Neuberger:1998wv}:
\bea
D_{ov}&=&\frac{1}{a}\bigg(1+A(\gamma_5A\gamma_5A)^{-\frac{1}{2}}\bigg)
\nn\\
&=&\frac{1}{a}\bigg(1+A(A^{\dagger}A)^{-\frac{1}{2}}\bigg).
\eea
Therefore, we can show that \cite{Neuberger:1998wv}:
\bea
aD_{ov}D_{ov}^{\dagger}&=&\frac{1}{a}\big(1+\gamma_5\mathrm{sign}(Q)\big)\big(1+\mathrm{sign}(Q)\gamma_5\big)
\nn\\
&=&\frac{1}{a}\big(1+\gamma_5\mathrm{sign}(Q)+\mathrm{sign}(Q)\gamma_5+1)
\nn\\
&=&D_{ov}+D_{ov}^{\dagger}.
\eea
The overlap construction transforms the non-chiral Dirac operator $D_W$ to a solution satisfying the Ginsparg-Wilson relation for establishing the lattice chiral symmetry.

\subsection{(1+1)D Hamiltonian Formulation}
\noindent
When considering the (1+1)D Hamiltonian, we only have one spatial direction on the Wilson-Dirac operator.
Therefore, one can show \cite{Hayata:2023zuk}
\bea
A^{\dagger}(s=0)A(s=0)=1.
\eea
Hence, it implies that the overlap operator is the same as the Wilson-Dirac operator for a particular choice of $s$ \cite{Hayata:2023zuk}, $s=0$,
\bea
D_{ov}(s=0)=\frac{1}{a}\big(1+A(s=0)\big)=D_W.
\eea
This result is exciting because it implies that the Wilson fermion also has the lattice chiral symmetry as in the overlap fermion.
Because the Wilson term generates the unwanted pole to the propagator, it implies that the lattice chiral symmetry allows non-zero physical mass on a finite lattice.

\subsection{Non-Hermitian Lattice Formulation}
\noindent
The most straightforward method of realizing the non-Hermitian lattice formulation is to adopt the one-sided lattice differences \cite{Stamatescu:1993ga}.
The lattice action adopting the forward finite-difference is \cite{Stamatescu:1993ga}:
\bea
S_{F1}&=&a\sum_{n=0}^{N-1}\bar{\psi}(n)\bigg(\gamma_1\frac{\psi(n+1)-\psi(n)}{a}+m\psi(n)\bigg)
\nn\\
&=&a\sum_{n_1, n_2; \alpha_1, \alpha_2}\bar{\psi}(n_1)_{\alpha_1}\big(D(n_1, n_2)_{\alpha_1, \alpha_2}
\nn\\
&&
+m\delta_{n_1, n_2}\delta_{\alpha_1, \alpha_2}\big)\psi(n_2)_{\alpha_2},
\eea
where
\bea
D(n_1, n_2)_{\alpha_1, \alpha_2}\equiv
(\gamma_1)_{\alpha_1, \alpha_2}\frac{\delta_{n_1+1, n_2}-\delta_{n_1, n_2}}{a}.
\eea
The lattice propagator in the infinite size limit is \cite{Guo:2021sjp}:
\bea
S_{11}(x)&\rightarrow&\Bigg\{\begin{array}{ll}
(1-ma)^{n-1}, & x> 0 \\
0, & x<0
\end{array};
\nn\\
S_{22}(x)&\rightarrow&\Bigg\{\begin{array}{ll}
0, & x> 0 \\
(1+ma)^{n-1}, & x<0
\end{array}.
\eea
The continuum limit ($ma, a/x\rightarrow 0$) shows the exact propagator as in the free Dirac fermion theory \cite{Guo:2021sjp}.
We only have one pole in the continuum limit \cite{Guo:2021sjp}.
Because we replace $i\sin(pa)$ with $\exp(ipa),$ the number of poles reduces by half compared to the naive fermion.
\\

\noindent
When $d>1$, the non-physical poles appear, but these poles do not provide the additional degrees of freedom when taking the continuum limit \cite{Stamatescu:1993ga}.
We aim to demonstrate such a result from the 2D case.
When considering the non-Hermitian lattice formulation, the Dirac matrix is represented as follows \cite{Stamatescu:1993ga}:
\bea
\bar{D}(p)&=&\frac{1}{a}\sum_{\mu=1}^4\gamma_{\mu}\big(\exp(ip_{\mu}a)-1\big)
\nn\\
&=&\frac{2i}{a}\sum_{\mu=1}^2\gamma_{\mu}e^{i\frac{p_{\mu}a}{2}}\sin\bigg(\frac{p_{\mu}a}{2}\bigg).
\eea
The inverse Dirac matrix is \cite{Stamatescu:1993ga}
\bea
\bar{D}^{-1}(p)=\frac{-\frac{i}{2a}\sum_{\mu=1}^2\gamma_{\mu}e^{i\frac{p_{\mu}a}{2}}\sin(\frac{p_{\mu}a}{2})}
{\frac{1}{a^2}\sum_{\nu=1}^2e^{ip_{\nu}a}\sin^2(\frac{p_{\nu}a}{2})}.
\eea
It is straightforward to find the following non-physical poles:
\bea
p_1a=-p_2a=\pm\frac{\pi}{2}.
\eea
However, the expansion of $\exp(ip_{\mu}a)\sin^2(p_{\mu}a/2)/a^2$ around these non-physical poles does not shift the mass pole because there is no quadratic term \cite{Stamatescu:1993ga}.
When the non-physical poles shift mass, the mass becomes infinite in the continuum limit \cite{Stamatescu:1993ga}.
Generalizing to the general $d$-dimension can also evade the fermion doubling \cite{Stamatescu:1993ga}.
\\

\noindent
The one-sided lattice difference breaks the hypercubic symmetry \cite{Stamatescu:1993ga}.
In general, loss of the hypercubic symmetry implies non-renormalizability \cite{Stamatescu:1993ga}.
We need to consider the quenched averaging, which imposes the hypercubic symmetry at the observable level \cite{Stamatescu:1993ga}
\bea
\overline{\langle{\cal O}\rangle}\equiv\frac{1}{2^d}\sum_{\epsilon_{\mu}=\pm 1}\langle{\cal O}(\epsilon_{\mu})\rangle_{\epsilon_{\mu}}.
\eea

\section{Numerical Algorithms}
\label{sec:5}
\noindent
We introduce the MC method: Metropolis and Hybrid Monte Carlo (HMC).
The HMC is the most efficient method for simulating the lattice field theory.
Since the lattice fermion field is represented as a Grassmann variable, we aim to improve the naive implementation of the Monte Carlo method by substituting the fermion field with a pseudo-fermion field (a bosonic field).
We also introduce the necessary numerical algorithms for generating random numbers following the Gaussian distribution and solving the linear equations.
Because the non-Hermitian lattice formulation loses the Hermiticity in the Lagrangian, we cannot naively apply the MC simulation to the lattice system.
This issue can be avoided when the partition function is non-negative \cite{Guo:2021sjp}.

\subsection{MC Method}
\noindent
We first introduce the basic idea of the MC method.
We then see how this method reaches the equilibrium through the detailed balance.
The numerical algorithm of the MC method is based on the principles introduced by Metropolis and HMC.

\subsubsection{Basic Idea}
\noindent
Systems with a significant physical degree of freedom are interesting in physics.
To investigate such systems, one should involve an evaluation of higher-dimensional integration.
The exact solution for the integral is generally unknown.
A straightforward numerical study like the Runge-Kutta method is also impossible.
When considering $N$ atoms at a finite temperature $T$, the integration must be evaluated at $10^{3N}$ points if each site takes ten different values.
Hence, it is impossible to do a large $N$ study.
The main idea of the MC simulation is not to evaluate the integrand at each point but to sample a point with the Boltzmann distribution in the integrand.
The MC method can estimate a multiple integral better than a direct numerical study.
\\

\noindent
Consider the integral
\bea
I=\int_0^1 dx\ f(x).
\eea
Choosing $N$ points randomly (with equal probability) in the interval $\lbrack 0, 1\rbrack$,
\bea
I\approx\frac{1}{N}\sum_{j=1}^Nf(x_j).
\eea
This method is known as simple sampling.
\\

\noindent
Now introduce a weight function $w(x)$ satisfying
\bea
\int_0^1dx\ w(x)=1.
\eea
The integral can be expressed in the following way:
\bea
I&=&\int_0^1dx\ w(x)\frac{f(x)}{w(x)}
=\int_0^1dy\ \frac{f\big(x(y)\big)}{w\big(x(y)\big)}
\nn\\
&=&\int_0^1dy\ F(y),
\eea
where
\bea
y(x)\equiv \int_0^x dt\ w(t).
\eea
Sampling $y$ uniformly is equivalent to sampling $x$ using the weight function $w(x)$.
Choosing more sampling at locations with a larger $f$ is more effective.
This technique is known as importance sampling.
It is also the main idea of the MC simulation.
\\

\noindent
Now, we give an example to show the weight function.
We choose that:
\bea
f(x)=\frac{1}{1+x^2}; \qquad I=\int_0^1 dx\ f(x);
\nn
\eea
\bea
w(x)=\frac{4-2x}{3}; \qquad y=\int_0^xdx\ w(x)=\frac{x(4-x)}{3};
\nn
\eea
\bea
x=2-\sqrt{4-3y}.
\eea
However, it is hard to find such a weight function in general.
Hence, we will introduce some algorithms for a more general study.

\subsubsection{Detailed Balance}
\noindent
The detailed balance is
\bea
\frac{P(\vec{y}\rightarrow\vec{x})}{P(\vec{x}\rightarrow\vec{y})}=\frac{w(\vec{x})}{w(\vec{y})},
\eea
in which the $P(\vec{x}\rightarrow\vec{y})$ is the probability of transition from the $\vec{x}$ to the $\vec{y}$.
The transition probability can be expressed as the product of the trial step probability from $\vec{x}$ to $\vec{y}$, denoted as $T(\vec{x}\rightarrow\vec{y})$, and the acceptance probability of that step, $A(\vec{x}\rightarrow\vec{y})$
\bea P(\vec{x}\rightarrow\vec{y})=T(\vec{x}\rightarrow\vec{y})\cdot A(\vec{x}\rightarrow\vec{y}).
\eea
Later, we will show that each algorithms have a probability distribution of $w(\vec{x})$ when a system goes to equilibrium.

\subsubsection{Metropolis}
\noindent
The Metropolis algorithm is a general scheme for producing random variables with a given probability distribution of an arbitrary form.
It only requires calculating a weight function at each step of updating.
We first introduce the algorithm and then show the detailed balance for the algorithm.
The algorithm of Metropolis for a system is:
\begin{itemize}
\item (a) The new configuration $\sigma_j^{\prime}$ is selected with an arbitrary probability distribution satisfying
\bea
T(\sigma_j\rightarrow\sigma_j^{\prime})=T(\sigma_j^{\prime}\rightarrow\sigma_j).
\eea
\item (b) Compute the change in energy
\bea
\Delta E=E(\sigma_j^{\prime})-E(\sigma_j).
\eea
\item (c) If
\bea
\Delta E\le 0,
\eea
the new configuration $\sigma_j^{\prime}$ is accepted, otherwise generate a random number $r\in (0, 1)$ with uniform deviate.
If
\bea
r\le\exp(-\beta\Delta E),
\eea
the new configuration is accepted.
This provides the
\bea
A(\sigma_j\rightarrow\sigma_j^{\prime})
=\min\bigg(1, \frac{\exp\big(-\beta E(\sigma^{\prime})\big)}{\exp\big(-\beta E(\sigma)\big)}\bigg).
\eea
\item (d) Repeat the same procedure for another configuration.
\end{itemize}

\noindent
The main limitation of the Metropolis algorithm is that the likelihood of accepting a new configuration is relatively low unless the new one closely resembles an old one.
This drawback leads to a strongly correlated configuration with a long correlation time, which creates a serious problem at a phase transition point.
This phenomenon is known as a critical slowdown.
\\

\noindent
In the Metropolis algorithm, it satisfies:
\bea
T(\vec{x}\rightarrow\vec{y})&=&T(\vec{y}\rightarrow\vec{x}); 
\nn\\
 A(\vec{x}\rightarrow\vec{y})
&=&\min\bigg(1, \frac{w(\vec{y})}{w(\vec{x})}\bigg).
\eea
At the equilibrium, the net number of points moving from the $\vec{x}$ to another point $\vec{y}$ in the next step is:
\bea
\delta\rho(\vec{x})=\rho_n(\vec{x})P(\vec{x}\rightarrow\vec{y})-\rho_n(\vec{y})P(\vec{y}\rightarrow\vec{x})=0, 
\nn\\
\eea
where $\rho_n(\vec{x})$ is the density of points at $\vec{x}$ after $n$ steps.
Hence, we obtain the detailed balance as follows:
\bea
\frac{\rho_n(\vec{x})}{\rho_n(\vec{y})}&=&
\frac{P(\vec{y}\rightarrow\vec{x})}{P(\vec{x}\rightarrow\vec{y})}
=\frac{T(\vec{y}\rightarrow\vec{x})A(\vec{y}\rightarrow\vec{x})}{T(\vec{x}\rightarrow\vec{y})A(\vec{x}\rightarrow\vec{y})}
=\frac{A(\vec{y}\rightarrow\vec{x})}{A(\vec{x}\rightarrow\vec{y})}
\nn\\
&=&\frac{\min\bigg(1,\frac{w(\vec{x})}{w(\vec{y})}\bigg)}{\min\bigg(1, \frac{w(\vec{y})}{w(\vec{x})}\bigg)}=\frac{w(\vec{x})}{w(\vec{y})}.
\nn\\
\eea

\subsubsection{HMC}
\noindent
To improve the drawback of Metropolis, we need an algorithm with the following properties:
\begin{itemize}
\item{(i) efficient computation;
}
\item{(ii) acceptance probability is high even for a large lattice;
}
\item{(iii) weak correlation between successive configurations.
}
\end{itemize}
The HMC is an algorithm for updating all fields simultaneously with a high enough probability of accepting efficiently.
Consider a field theory with the action $S(\phi)$, where $\phi$ denotes a generic field variable:
\begin{itemize}
\item{
1. Initial configuration $\phi$.
}
\item{2. Generate momenta $\pi$ with Gaussian distribution
\bea
P_G(\pi)\propto \exp(-\pi^2/2).
\eea
}
\item{3. Molecular dynamics step (time length=1):
\bea
&&
\dot{\phi}=\frac{\partial H}{\partial\pi}, \qquad \dot{\pi}=-\frac{\partial H}{\partial\phi}=-\frac{\partial S}{\partial\phi}, 
\nn\\
&&
H(\phi, \pi)=\frac{\pi^2}{2}+S(\phi).
\eea
}
\item{4. $P_A(\{\phi, \pi\}\rightarrow\{\phi^{\prime}, \pi^{\prime}\})=\min\big(1, \exp(-\Delta H)\big)$, where
\bea
\Delta H=H(\phi^{\prime}, \pi^{\prime})-H(\phi, \pi).
\eea
}
\item{
5. Go to 2.
}
\end{itemize}
The leap-frog algorithm for molecular dynamics (Molecular dynamics step) is:
\begin{itemize}
\item{
1. Initial half-step for the $\pi$:
\bea
\phi(0)=\phi; \qquad \pi\bigg(\frac{\epsilon}{2}\bigg)=\pi(0)-\frac{\epsilon}{2}\frac{\partial S(\phi)}{\partial\phi(0)},
\eea
where $\epsilon$ is the discrete time step.
}
\item{
2. Intermediate $N-1$ steps ($\tau=n\epsilon, n=1, 2, \cdots, N-1$):
\bea
\phi(\tau)&=&\phi(\tau-\epsilon)+\epsilon\cdot\pi\bigg(\tau-\frac{\epsilon}{2}\bigg);
\nn\\
\pi\bigg(\tau+\frac{\epsilon}{2}\bigg)&=&\pi\bigg(\tau-\frac{\epsilon}{2}\bigg)-\epsilon\frac{\partial S(\phi)}{\partial\phi(\tau)}.
\eea
}
\item{
3. Final step for the $\phi$ (half-step for the $\pi$):
\bea
\phi^{\prime}&=&\phi(1)=\phi(1-\epsilon)+\epsilon\cdot\pi\bigg(1-\frac{\epsilon}{2}\bigg);
\nn\\
\pi^{\prime}&=&\pi(1)=\pi\bigg(1-\frac{\epsilon}{2}\bigg)-\frac{\epsilon}{2}\frac{\partial S(\phi)}{\partial\phi(1)}.
\eea
}
\end{itemize}
The leap-frog algorithm has the reversibility
\bea
\{\phi^{\prime}, -\pi^{\prime}\}\rightarrow\{\phi,-\pi\}.
\eea

\noindent
The detailed balance is demonstrated as follows:
\begin{itemize}
\item{
1. $P_G=c\cdot \exp(-\pi^2/2)$ and $\int d\pi\ P_G(\pi)=1$.
}
\item{
2. Reversibility is the molecular dynamics:
\bea
\{\phi, \pi\}\rightarrow\{\phi^{\prime}, \pi^{\prime}\}&\Rightarrow&\{\phi^{\prime}, -\pi^{\prime}\}\rightarrow\{\phi,-\pi\};
\nn\\
P_M(\{\phi, \pi\}\rightarrow\{\phi^{\prime}, \pi^{\prime}\})&=&P_M(\{\phi^{\prime}, -\pi^{\prime}\}\rightarrow\{\phi, -\pi\}).
\nn\\
\eea
}
\item{
3. The Metropolis test
\bea
P_A(\{\phi, \pi\}\rightarrow\{\phi^{\prime}, \pi^{\prime}\})=\min\big(1, \exp(-\Delta H)\big),
\nn\\
\eea
where
\bea
\Delta H&=&H(\phi^{\prime}, \pi^{\prime})-H(\phi, \pi); 
\nn\\
H(\phi, \pi)&=&\frac{\pi^2}{2}+S(\phi).
\eea
Then we have:
\bea
P(\phi\rightarrow\phi^{\prime})&=&\int d\pi d\pi^{\prime}\ P_G(\pi)P_M(\{\phi, \pi\}
\nn\\
&\rightarrow&\{\phi^{\prime}, \pi^{\prime}\})
P_A(\{\phi, \pi\}\rightarrow\{\phi^{\prime}, \pi^{\prime}\}),
\nn\\
P_{\mathrm{eq}}(\phi)&\propto& e^{-S(\phi)},
\nn\\
\int d\phi\ P_{\mathrm{eq}}(\phi)&=&1.
\eea
}
\item{
Finally, we show that:
\bea
&&P_{\mathrm{eq}}(\phi)P(\phi\rightarrow\phi^{\prime})
\nn\\
&=&\int d\pi d\pi^{\prime}\ \exp\big(-H(\phi, \pi)\big)P_M(\{\phi, \pi\}\rightarrow\{\phi^{\prime}, \pi^{\prime}\})
\nn\\
&&\times
P_A(\{\phi, \pi\}\rightarrow\{\phi^{\prime}, \pi^{\prime}\})
\nn\\
&=&\int d\pi d\pi^{\prime}\ \exp\big(-H(\phi^{\prime}, \pi^{\prime})\big)
\nn\\
&&\times
P_M(\{\phi^{\prime}, -\pi^{\prime}\}
\rightarrow\{\phi, -\pi\})P_A(\{\phi^{\prime}, \pi^{\prime}\}\rightarrow\{\phi, \pi\})
\nn\\
&=&\int d(-\pi)d(-\pi^{\prime})\ \exp\big(-H(\phi^{\prime}, -\pi^{\prime})\big)
\nn\\
&&\times
P_M(\{\phi^{\prime}, -\pi^{\prime}\}\rightarrow\{\phi, -\pi\})
P_A(\{\phi^{\prime}, -\pi^{\prime}\}\rightarrow\{\phi, -\pi\})
\nn\\
&=&P_{\mathrm{eq}}(\phi^{\prime})P(\phi^{\prime}\rightarrow\phi),
\eea
in which we used:
\bea
&&\exp\big(-H(\phi, \pi)\big)P_A(\{\phi, \pi\}\rightarrow\{\phi^{\prime}, \pi^{\prime}\})
\nn\\
&=&\exp\big(-H(\phi, \pi)\big)\times\min\bigg(1, \exp\big(-H(\phi^{\prime}, \pi^{\prime})+H(\phi, \pi)\big)\bigg)
\nn\\
&=&\bigg\{\begin{array}{ll}
\exp\big(-H(\phi, \pi)\big), & H(\phi^{\prime}, \pi^{\prime})\le H(\phi, \pi) \\
\exp\big(-H(\phi^{\prime}, \pi^{\prime}\big), & H(\phi^{\prime}, \pi^{\prime})> H(\phi, \pi)
\end{array} ;
\eea
\bea
&&\exp\big(-H(\phi^{\prime}, \pi^{\prime})\big)P_A(\{\phi^{\prime}, \pi^{\prime}\}\rightarrow\{\phi, \pi\})
\nn\\
&=&\exp\big(-H(\phi^{\prime}, \pi^{\prime})\big)\times\min\bigg(1, \exp\big(H(\phi^{\prime}, \pi^{\prime})-H(\phi, \pi)\big)\bigg)
\nn\\
&=&\bigg\{\begin{array}{ll}
\exp\big(-H(\phi^{\prime}, \pi^{\prime})\big), & H(\phi, \pi)< H(\phi^{\prime}, \pi^{\prime}) \\
\exp\big(-H(\phi, \pi\big), & H(\phi, \pi)\ge H(\phi^{\prime}, \pi^{\prime})
\end{array};
\eea
\bea
&&\exp\big(-H(\phi, \pi)\big)P_A(\{\phi, \pi\}\rightarrow\{\phi^{\prime}, \pi^{\prime}\})
\nn\\
&=&\exp\big(-H(\phi^{\prime}, \pi^{\prime})\big)P_A(\{\phi^{\prime}, \pi^{\prime}\}\rightarrow\{\phi, \pi\})
\nn\\
\eea
in the second equality.
}
\end{itemize}

\subsection{Pseudofermion Method}
\noindent
We introduce the pseudofermion method to implement the MC method to simulate the lattice fermion fields.
Here, we demonstrate this method from the following lattice theory
\bea
L_L=L_{\phi}+\sum_j\bar{\psi}_jD_{\phi}\psi_j,
\eea
in which $L_{\phi}$ is the Lagrangian for a lattice scalar field theory.
The label $\phi$ in the $D_{\phi}$ indicates that the fermion field interacts with the scalar field.
In general, integrating out fermion fields would give a non-positive definite determinant.
The application of the MC simulation to a non-positive definite matrix cannot have the importance of sampling.
We introduce the pseudofermion method for the case with the positive-definite determinant.
\\

\noindent
When considering two fermion fields as an example
\bea
&&
\int {\cal D}\bar{\psi}{\cal D}\psi\ \exp(-\bar{\psi}_1D\psi_1-\bar{\psi}_2D\psi_2)
\nn\\
&\sim&
\int {\cal D}\phi_{f, R}{\cal D}\phi_{f, I}\ \exp\big(-\phi_f^{\dagger}(DD^{\dagger})^{-1}\phi_f\big),
\eea
where
\bea
\phi_f\equiv \phi_{f, R}+i\phi_{f, I},
\eea
the $DD^{\dagger}$ is a positive-definite matrix.
We assume that the Dirac matrix satisfies the $\gamma_5$-Hermiticity.
Therefore, the determinant of $D$ is equivalent to the determinant of $D^{\dagger}$.
The $\phi_f$ is the bosonic field, and it is called the pseudofermionic field.
We do not have Grassmann variables in the path integration, but we meet the non-local action.
\\

\noindent
The first step of implementation is to introduce the momenta $\pi_j^*$, $\pi_j$, and $P_l$ canonically conjugate to $\phi^*_{f, j}$, $\phi_{f, j}$, and $\phi_l$ respectively.
The Hamiltonian  is
\bea
H=
\frac{1}{2}\sum_lP_l^2+\sum_j\pi_j^*\pi_j+S_{\phi}(\phi)+S_{pf}(\phi, \phi_f, \phi_f^*),
\nn\\
\eea
where
\bea
S_{pf}\equiv\sum_{j,k}\phi_{f, j}^*(Q^{-1})_{jk}\phi_{f, k};\ Q\equiv DD^{\dagger}.
\eea
The partition function is represented as follows
\bea
Z\equiv\int D\phi D\phi_f D\phi_f^* DP D\pi D\pi^*\ e^{-H}.
\eea
The following equations describe the molecular dynamics step:
\bea
\dot{\phi}_l&=&\frac{\partial H}{\partial P_l}=P_l;
\nn\\
\dot{P}_l&=&-\frac{\partial H}{\partial\phi_l}=-\frac{\partial S_{\phi}}{\partial\phi_l}
-\sum_{j, k}\phi_{f, j}^*\frac{\partial (Q^{-1})_{jk}}{\partial\phi_l}\phi_{f,k};
\nn\\
\dot{\phi}_{f, j}&=&\frac{\partial H}{\partial \pi^*_j}=\pi_j;
\nn\\
\dot{\pi}_j&=&-\frac{\partial H}{\partial\phi_{f, j}^*}=-\sum_k(Q^{-1})_{jk}\phi_{f, k}.
\eea
The equation above can be simplified as follows:
\bea
\eta_j&=&\sum_{k}(Q^{-1})_{jk}\phi_{f, k};
\nn\\
\dot{\phi}_{f, j}&=&\pi_l;
\nn\\
\dot{\pi}_j&=&-\eta_j;
\nn\\
\dot{\phi}_l&=&P_l;
\\
\dot{P}_l&=&-\frac{\partial S_{\phi}}{\partial\phi_l}
+\sum_{j, k}\eta^*_j\frac{\partial Q_{jk}}{\partial\phi_l}\eta_k,
\eea
in which we used:
\bea
0&=&\frac{\partial}{\partial{\phi_l}}(QQ^{-1})
=\bigg(\frac{\partial}{\partial{\phi_l}}Q\bigg)Q^{-1}
+Q\bigg(\frac{\partial}{\partial{\phi_l}}Q^{-1}\bigg);
\nn\\
\frac{\partial}{\partial{\phi_l}}Q^{-1}&=&-Q^{-1}\bigg(\frac{\partial}{\partial{\phi_l}}Q\bigg)Q^{-1}
\eea
To perform the molecular dynamics practically, we rewrite the $S_{pf}$ as below:
\bea
S_{pf}
=\sum_{j,k}\phi^*_{f, j}\big((D^{\dagger})^{-1}D^{-1}\big)_{jk}\phi_{f, k}
=\sum_j\xi_j^*\xi_j,
\eea
where
\bea
\xi_j\equiv (D^{-1})_{jk}\phi_{f, k}.
\eea
\\

\noindent
Now, we can generate configurations in the variable $\xi$, which follow the Gaussian distribution.
The value of $\phi_f$ can be calculated using the fixed configuration of the scalar field $\phi$,
\bea
\phi_f=D\xi.
\eea
An ensemble of configurations for the pseudofermionic variables with the distribution $\exp(-\phi_f^{\dagger}Q^{-1}\phi_f)$.
When we generate the $\xi$ with the Gaussian distribution, the simulation is unnecessary to generate $\pi$, and the pseudofermion field becomes a background field without the evolving of molecular dynamics.
Indeed, this method is more practical.
The algorithm of HMC is as in the following:
\begin{itemize}
\item{
1. Choose an initial configuration for the scalar field $\phi$.
}
\item{
2. Choose $P_l$ from a Gaussian ensemble as
\bea
\exp\bigg(-\frac{1}{2}\sum_lP_l^2\bigg).
\eea
}
\item{
3. Choose $\xi$ from the Gaussian distribution as $\exp(\xi^{\dagger}\xi)$.
}
\item{
4. Calculate
\bea
\phi_f=D\xi.
\eea
}
\item{
5. Allow the $\phi_l$ and its canonical momenta $P_l$ to evolve according to:
\bea
\dot{\phi}_l&=&P_l;
\\
\dot{P}_l&=&-\frac{\partial S_{\phi}}{\partial\phi_l}
+\sum_{j, k}\eta^*_j\frac{\partial Q_{jk}}{\partial\phi_l}\eta_k,
\eea
where $\eta=Q^{-1}\phi_f$ with the fixed configuration of scalar field $\phi_f$.
}
\item{
6. Accept the new configuration $(\tilde{\phi}, \tilde{P})$ with the probability
\bea
\mathrm{min}\bigg(1, \frac{\exp\big(-\tilde{H}[\tilde{\phi}, \tilde{P}]\big)}
{\exp\big(-\tilde{H}[\phi, P]\big)}\bigg),
\eea
where
\bea
\tilde{H}[\phi, P]=\frac{1}{2}\sum_lP_l^2
+S_{\phi}[\phi]
+\sum_j\xi^*_j\xi_j.
\eea
}
\item{
7. Save the configuration, whether new or old, based on the results of the Metropolis test.
}
\item{
8. Return to 2.
}
\end{itemize}

\subsection{Box-Muller Method}
\noindent
We first give the algorithm for generating a random number with an arbitrary probability distribution and then apply it to the Gaussian distribution.
A uniformly distributed random number is:
\bea
x\in\lbrack 0, 1\rbrack; \qquad x\equiv\int_{-\infty}^ydy\ P(y)\equiv F(y),
\eea
where $P(y)$ is a probability distribution.
It is easy to know:
\bea
F(-\infty)=0;\qquad F(\infty)=1.
\eea
The inverse $F^{-1}$  gives
\bea
F^{-1}(x)=y(x).
\eea
Therefore, the variable $x$ with a uniform distribution is equivalent to the variable $y$ with a probability distribution $P(y)$:
\bea
\int dx=\int dy\frac{dx}{dy}=\int dy\ P(y).
\eea
\\

\noindent
The random numbers with Gaussian distribution
\bea
p(x)=\frac{1}{\sqrt{2\pi}}\exp\bigg(-\frac{x^2}{2}\bigg)
\eea
can be generated by the Box-Muller method. Consider points $(x_1, x_2)$ in a plane where $x_1$ and $x_2$ both have Gaussian distribution:
\bea
p(x_1)&=&\frac{1}{\sqrt{2\pi}}\exp\bigg(-\frac{x_1^2}{2}\bigg), 
\nn\\
p(x_2)&=&\frac{1}{\sqrt{2\pi}}\exp\bigg(-\frac{x_2^2}{2}\bigg).
\eea
Then the number of points in the $dx_1dx_2$ is proportional to:
\bea
p(x_1)p(x_2)dx_1dx_2=\frac{1}{2\pi}e^{-\frac{x_1^2+x_2^2}{2}}dx_1dx_2=\frac{1}{2\pi}e^{-u}dud\theta,
\nn\\
\eea
where
\bea
x_1\equiv\sqrt{2u}\cos(\theta), \qquad x_2=\sqrt{2u}\sin(\theta).
\eea
The variable $u\in (0, \infty)$ has an exponential distribution, and $\theta$ uniformly between 0 and $2\pi$, and then both will
have the desired Gaussian distribution.
Hence, we have the following:
\bea
x&=&F(R)=\frac{1}{2\pi}\int_0^{2\pi} d\theta\int_0^{\frac{R^2}{2}} du\ e^{-u}
\nn\\
&=&1-\exp\bigg(-\frac{R^2}{2}\bigg).
\eea
Because:
\bea
F(0)=0;\qquad F(\infty)=1,
\eea
the $x$ can be uniformly distributed on the unit interval $(0, 1)$.
Therefore, we obtain:
\bea
R(x)=\sqrt{-2\ln(1-x)}\equiv\sqrt{-2\ln R_1}.
\eea
To generate the random variables, we choose:
\bea
R(x)\equiv\sqrt{-2\ln R_1};\qquad \theta\equiv 2\pi R_2.
\eea
$R_1$ and $R_2$ are independent random numbers with a uniform distribution on the unit interval $(0, 1)$.
Hence using:
\bea
x_1&=&\sqrt{-2\ln R_1}\cos(2\pi R_2); 
\nn\\
x_2&=&\sqrt{-2\ln R_1}\sin(2\pi R_2)
\eea
shows the equivalence
\bea
\int d R_1dR_2=\frac{1}{2\pi}\int dx_1dx_2\ \exp\bigg(-\frac{x_1^2+x_2^2}{2}\bigg).
\nn\\
\eea

\subsection{Conjugate Gradient Method}
\noindent
The conjugate gradient method is an algorithm for obtaining $x$ from the matrix equation
\bea
Ax=b
\eea
with a given positive-semidefinite matrix $A$ and $b$.
The idea is to minimize
\bea
S(x)=\frac{1}{2}|b-Ax|^2.
\eea
Assuming
\bea
x_{k+1}\equiv x_k+\lambda_kp_k,
\eea
the minimization of the $S(x)$ gives the condition:
\bea
0
&=&\frac{dS(x_{k+1})}{d\lambda_k}
=\frac{1}{2}\frac{d}{d\lambda_k}|b-Ax_k-\lambda_kAp_k|^2
\nn\\
&\equiv&
\frac{1}{2}\frac{d}{d\lambda_k} (b-Ax_k-\lambda_kAp_k, b-Ax_k-\lambda_kAp_k)
\nn\\
&=&
\lambda_k(Ap_k, Ap_k)-(Ap_k, b-Ax_k),
\eea
which implies:
\bea
\lambda_k=\frac{(Ap_k, b-Ax_k)}{(Ap_k, Ap_k)}
=
\frac{(p_k, s_k)}{(Ap_k, Ap_k)}.
\eea
The $s_k$ is defined by:
\bea
s_k\equiv A^{\dagger}r_k;\ r_k\equiv b-Ax_k.
\eea
For an $n\times n$ matrix $A$,
\bea
x=x_0+\sum_{j=0}^{n-1}\lambda_jp_j,
\eea
in which $p_0$, $p_1$, $\cdots$, $p_{n-1}$ are linearly independent for $A^{\dagger}A$
\bea
(p_j, A^{\dagger}A p_k)=0
\eea
for $j\neq k$.
Now we assume
\bea
x_{j+1}\equiv x_0+\sum_{k=0}^{j}\lambda_kp_j.
\eea
Therefore, this implies:
\bea
x_{k+1}=x_{j+1}+\sum_{l=j+1}^k\lambda_lp_l;
\nn
\eea
\bea
s_{k+1}&=&A^{\dagger}(b-Ax_{k+1})
=A^{\dagger}\bigg(b-Ax_{j+1}-\sum_{l=j+1}^k\lambda_lAp_l\bigg)
\nn\\
&=&s_{j+1}-\sum_{l=j+1}^k\lambda_lA^{\dagger}Ap_l;
\nn
\eea
\bea
(p_j, s_{k+1})&=&
(p_j, s_{j+1})-\sum_{l=j+1}^k\lambda_l(p_j, A^{\dagger}Ap_l)
\nn\\
&=&(p_j, s_{j+1})
\nn\\
&=&\big(p_j, A^{\dagger}(b-Ax_j-\lambda_jAp_j)\big)
\nn\\
&=&(p_j, s_j)
-\lambda_j(Ap_j, Ap_j)
\nn\\
&=&
0.
\eea
\\

\noindent
Now we choose
\bea
p_k=s_k+\mu_{k-1}p_{k-1},
\eea
which leads to the following conditions:
\bea
0&=&(p_{k-1}, A^{\dagger}Ap_k)
\nn\\
&=&(p_{k-1}, A^{\dagger}As_k)+\mu_{k-1}(p_{k-1}, A^{\dagger}Ap_{k-1}).
\eea
Therefore, we obtain
\bea
\mu_{k-1}=-\frac{(p_{k-1}, A^{\dagger}As_k)}{(A p_{k-1}, A p_{k-1})}.
\eea
\\

\noindent
We first assume
\bea
(s_j, s_l)=0,\ j,l\le k.
\eea
By using:
\bea
0&=&(p_{k-1}, s_k)=(s_{k-1}+\mu_{k-2}p_{k-2}, s_k)
\nn\\
&=&\mu_{k-2}(p_{k-2}, s_k),
\eea
the derivation can be done iterative, and it shows
\bea
(p_{j-1}, s_k)=0,\ j\le k.
\eea
Hence, we obtain the:
\bea
0&=&(p_j, s_{k+1})=(s_{j}+\mu_{j-1}p_{j-1}, s_{k+1})
\nn\\
&=&(s_j, s_{k+1}),
\eea
which leads
\bea
(s_j, s_l)=0,\ j,l\le k+1,\ j\neq l.
\eea
We induce the above result from the assumption.
Hence we prove
\bea
(s_j, s_l)=0,\ j\neq l.
\eea
Because the $s_j$ are linearly independent, where
\bea
j=0, 1, \cdots, n-1,
\eea
it guarantees:
\bea
s_n=A^{\dagger}(b-Ax_n)=0,
\eea
which implies
\bea
b-Ax_n=0.
\eea
Hence, linear independence can guarantee finding a solution with suitable accuracy.
\\

\noindent
In the final, we show that the choice also guarantees
\bea
(p_j, A^{\dagger}Ap_l)=0,\ j\neq l.
\eea
We first obtain the below result:
\bea
(p_j, A^{\dagger}Ap_{l+1})
=
(p_j, A^{\dagger}As_{l+1})
+\mu_{l}(p_j, A^{\dagger}Ap_{l});
\nn\\
\eea
\bea
(p_j, A^{\dagger}As_{l+1})
&=&\frac{1}{\lambda_j}(x_{j+1}-x_j, A^{\dagger}As_{l+1})
\nn\\
&=&-\frac{1}{\lambda_j}(s_{j+1}-s_j, s_{l+1})
\nn\\
&=&
0,
\eea
in which we used:
\bea
A^{\dagger}A(x_{j+1}-x_j)
&=&A^{\dagger}\big(-(b-Ax_{j+1})+(b-Ax_j)\big)
\nn\\
&=&A^{\dagger}(-r_{j+1}+r_j)
\nn\\
&=&-s_{j+1}+s_j
\nn\\
\eea
in the second equality;
\bea
(p_j, A^{\dagger}Ap_{l+1})
=\mu_{l}(p_j, A^{\dagger}Ap_{l}),\ j\neq l.
\eea
We first assume
\bea
(p_j, A^{\dagger}Ap_{l})=0,\ j, l\le k,\ j\neq l.
\eea
Then this leads
\bea
(p_j, A^{\dagger}Ap_{l})=0,\ j, l\le k+1,\ j\neq l.
\eea
The induction proves
\bea
(p_j, A^{\dagger}Ap_l)=0,\ j\neq l.
\eea
\\

\noindent
Based on the above discussion, the algorithm of conjugate gradient is:
\begin{itemize}
\item{Choosing an initial $x_0$ and then calculating
\bea
r_0=b-Ax_0, \qquad p_0=s_0=A^{\dagger}r_0.
\nn
\eea
\item{Repeating the below procedure for $k=0, 1, \cdots$:
\bea
\lambda_k&=&\frac{(s_k, p_k)}{(Ap_k, Ap_k)};
\nn\\
x_{k+1}&=&x_k+\lambda_kp_k;
\nn\\
r_{k+1}&=&b-Ax_{k+1};
\nn\\
\mathrm{If} &|r_{k+1}|&<\epsilon, \mathrm{then\ stop};
\nn\\
s_{k+1}&=&A^{\dagger} r_{k+1};
\nn\\
\mu_k&=&-\frac{(p_k, A^{\dagger}As_{k+1})}{(Ap_k, Ap_k)};
\nn\\
p_{k+1}&=&s_{k+1}+\mu_kp_k.
\nn
\eea
}
}
\end{itemize}
Indeed, the above algorithm has unnecessary matrix calculation, but it is more intuitive.
We will show some valuable identities for increasing the speed of the algorithm.
\\

\noindent
When we apply the conjugate gradient to solving a generic matrix, we solve the following positive-definite matrix $A^{\dagger}A$,
\bea
A^{\dagger}Ax\equiv Px=A^{\dagger}b\equiv\Phi.
\eea
Because the only difference is a multiplication of the $A^{\dagger}$, we still choose the minimization of the $S(x)$.
Now we define:
\bea
R_k\equiv s_k= A^{\dagger}(b-Ax_k)=\Phi-Px_k.
\eea
Therefore, this gives:
\bea
(p_k, s_k)&=&(p_k, R_k)=(R_k+\mu_{k-1}p_{K-1}, R_k)
\nn\\
&=&(R_k, R_k)+\mu_{k-1}(p_{k-1}, R_k).
\eea
Here we choose
\bea
p_k=R_k+\mu_{k-1}p_{k-1}.
\eea
We also have:
\bea
0=(p_{k-1}, s_k)
=(p_{k-1}, R_k).
\eea
Hence, we obtain the:
\bea
(p_k, s_k)=(R_k, R_k); \qquad \lambda_k=\frac{(R_k, R_k)}{(p_k, P p_k)}.
\eea
Due to the fact:
\bea
R_{k+1}-R_{k}=P(x_k-x_{k+1})=-\lambda_kPp_k,
\eea
it gives:
\bea
&&
(R_{k+1}, R_{k+1})
\nn\\
&=&(R_{k+1}, R_k)-\lambda_k(R_{k+1}, Pp_K)
\nn\\
&=&(R_{k+1}, R_k)-\frac{(R_k, R_k)}{(P_k, Pp_k)}(R_{k+1}, Pp_k).
\eea
Hence we obtain
\bea
(R_{k+1}, R_{k+1})=-\frac{(R_k, R_k)}{(P_k, Pp_k)}(R_{k+1}, Pp_k)
\eea
from
\bea
0=(s_{k+1}, s_k)=(R_{k+1}, R_k).
\eea
The $\mu_k$ is
\bea
\mu_k=\frac{(R_{k+1}, R_{k+1})}{(R_k, R_k)}.
\eea
The algorithm is:
\begin{itemize}
\item{Choosing an initial $x_0$ and then calculating:
\bea
R_0=\Phi-Px_0,\qquad p_0=R_0, \qquad k=0.
\nn
\eea
}
\item{Repeating the following procedure for $k=0, 1, \cdots$:
\bea
\lambda_k&=&\frac{(R_k, R_k)}{(p_k, Pp_k)};
\nn\\
R_{k+1}&=&R_{k}-\lambda_kPp_k;
\nn\\
\mathrm{If} &|R_{k+1}|&<\epsilon|\Phi|, \mathrm{then\ stop};
\nn\\
\mu_k&=&\frac{(R_{k+1}, R_{k+1})}{(R_k, R_k)};
\nn\\
x_{k+1}&=&x_k+\lambda_kp_k;
\nn\\
p_{k+1}&=&R_{k+1}+\mu_{k}p_k.
\nn
\eea
}
\end{itemize}
This algorithm is more efficient.

\subsection{Non-Hermitian Lattice Formulation}
\noindent
Applying the HMC method to the non-Hermitian lattice formulation is problematic because it loses the Hermiticity.
Implementing the Monte Carlo method only requires the non-negative partition function.
We demonstrate our simulation from 1D Dirac fermion fields with a degenerate mass for two flavors \cite{Guo:2021sjp}
\bea
&&
S_{\mathrm{FD}}
\nn\\
&=&a\sum_{n=0}^{N-1}\bigg(\bar{\psi}_1(n)\big(D(n)+m_{\mathrm{F}}\big)\psi_1(n)
\nn\\
&&
+
\bar{\psi}_2(n)\big(-D^{\dagger}(n)+m_{\mathrm{F}}\big)\psi_2(n)\bigg).
\eea
The forward finite-difference defines the derivative operator of $\psi_1$, and the derivative operator of another field $\psi_2$ is defined using the backward finite-difference.
After integrating out the fermion fields, we obtain a non-negative partition function \cite{Guo:2021sjp}:
\bea
&&
\det(D+m)\det(-D^{\dagger}+m_{\mathrm{F}})
\nn\\
&=&\det(D+m_{\mathrm{F}})\det\big(\gamma_5(-D^{\dagger}+m_{\mathrm{F}})\gamma_5\big)
\nn\\
&=&|\det(D+m_{\mathrm{F}})|^2.
\eea
We use
\bea
\gamma_5D+D\gamma_5=0
\eea
in the second equality.
We then introduce the pseudo-fermion field (bosonic field $\phi_f$) to rewrite the partition function as in the following
\bea
&&
\int {\cal D}\bar{\psi}{\cal D}\psi\ \exp(-S_{\mathrm{FD}})
\nn\\
&\sim&
\int {\cal D}\phi_{f, {\mathrm{R}}}{\cal D}\phi_{f, \mathrm{I}}\ \exp\big(-\phi_f^{\dagger}
\nn\\
&&\times
\big((D+m_{\mathrm{F}})(D^{\dagger}+m_{\mathrm{F}})\big)^{-1}\phi_f\big),
\eea
where
\bea
\phi_f\equiv\phi_{f, \mathrm{R}}+i\phi_{f, \mathrm{I}}.
\eea
Although it is a non-Hermitian field theory, the partition function is real-valued.
Hence, we can implement the HMC method to compute the observables in free Dirac fermions \cite{Guo:2021sjp}.
This method can have a similar extension to the even flavors and also the interacting theory \cite{Guo:2024jqt}.
\\

\noindent 
In the free theory, the non-Hermitian lattice formulation does not suffer from the renormalizability issue.
The initial simulation results from the non-Hermitian lattice formulation of the interacting 2D GNY model with the confirmation of the resummation techniques \cite{Guo:2024jqt} are valuable for exploring this issue.
The action of the GNY model is
\bea
S_{\mathrm{GNY}}\lbrack\bar{\psi}, \psi, \phi\rbrack
=
\int d^2x\ \bigg(\bar{\psi}(\fsl{\partial}+m_{\mathrm{F}}+\phi)\psi-\phi\Box\phi+\frac{1}{2g^2}\phi^2\bigg).
\eea
In the action of the GNY model, the term $\phi\Box\phi$ represents the kinetic term for the scalar field, while the term $\phi^2/(2g^2)$ denotes its self-interaction.
The coupling constant $g^2$ is positive.
The GNY model has the essential property of asymptotic safety \cite{Guo:2024jqt,Thirring:1958in,Skyrme:1958vn,Skyrme:1961vr,Coleman:1974bu,Mandelstam:1975hb,Buscher:1987sk,Buscher:1987qj,Burgess:1993np,Kovacs:2014fwa}, which helps study the interacting theory on a lattice \cite{Aizenman:1981du,Luscher:1987ay}.
      
\section{Chiral Anomaly and Index Theorem}
\label{sec:6}
\noindent
The chiral anomaly can be introduced through the chiral transformation applied to the measure \cite{Fujikawa:1979ay}.
This method can also be implemented on a lattice \cite{Fujikawa:1998if,Suzuki:1998yz}.
The chiral anomaly is relevant to the index theorem \cite{Atiyah:1968mp,Atiyah:1970ws,Atiyah:1971rm} or the topological charge \cite{Fujikawa:1979ay}.
Hence, we can apply the technique to realize the index theorem on a lattice \cite{Fujikawa:1998if,Suzuki:1998yz}.
However, the chiral anomaly on a lattice indicates topologically nontrivial background gauge fields, even if the topological charge is zero \cite{Chiu:2001bg,Chiu:2001ja}.
Hence, the non-trivial topological charge should rely on a non-trivially infinite lattice size limit \cite{Chiu:2001bg,Chiu:2001ja}.
We provide one solution of the Ginsparge-Wilson relation to explain such a situation \cite{Ginsparg:1981bj,Chiu:2001bg,Chiu:2001ja}.

\subsection{Chiral Anomaly}
\noindent
Consider the functional integral for massless Dirac fermions coupled to a gauge field $A_{\mu}$,
\bea
Z=\int [d\psi d\bar{\psi}]\ e^{-S(\psi, \bar{\psi})},
\eea
where
\bea
S(\psi, \bar{\psi})=\int d^4x\ \bar{\psi}\gamma_{\mu}(\partial_{\mu}+A_{\mu})\psi.
\eea
This classical action has chiral symmetry.
We now consider a change of variables
\bea
\tilde{\psi}=e^{i\gamma_5\alpha(x)}\psi, \ \bar{\tilde{\psi}}=\bar{\psi} e^{i\gamma_5\alpha(x)},
\eea
where $\alpha(x)$ is defined as an arbitrary function of the coordinates.
Because it is just a change of variables, the partition function $Z$ is unchanged \cite{Fujikawa:1979ay}:
\bea
Z
&=&\int [d\tilde{\psi} d\bar{\tilde{\psi}}]\ e^{-S(\tilde{\psi}, \bar{\tilde{\psi}})}
\nn\\
&=&\int [d\psi d\bar{\psi}]\ \mathrm{det}(e^{2i\gamma_5\alpha})e^{-S(\psi, \bar{\psi})}
\nn\\
&&\times
\exp\bigg(-\int d^4x\ \alpha\partial_{\mu}J_{\mu}\bigg)
\nn\\
&=&\int [d\psi d\bar{\psi}]\ e^{-S(\psi, \bar{\psi})}
\nn\\
&&\times
\exp\bigg(2i\mathrm{Tr}(\gamma_5\alpha)-\int d^4x\ \alpha\partial_{\mu}J_{\mu}\bigg),
\eea
where
\bea
J_{\mu}=i\bar{\psi}\gamma_{\mu}\gamma_5\psi.
\eea
The trace operation is over the infinite-dimensional spaces of eigenmodes and the Dirac matrices.
Since the measure is not invariant under chiral transformations, chiral symmetry is broken due to quantum effects \cite{Fujikawa:1979ay}.

\subsection{Topological Charge}
\noindent
After performing the trace operation $\mathrm{Tr}(\gamma_5)$, we can discover that the chiral anomaly can be relevant to the index theorem or the topological charge.
The non-zero eigenvalues and their corresponding eigenmodes are paired, meaning they will effectively cancel each other out.
We only need to be concerned about the zero eigenvalues.
Therefore, we can introduce the regulator $\Lambda$ and extract the result when $\Lambda\rightarrow\infty$,
\bea
&&
\mathrm{Tr}(\gamma_5)
\nn\\
&=&
\lim_{\Lambda^2\rightarrow\infty}\mathrm{Tr}\bigg(\gamma_5 e^{\frac{\big(\gamma_{\mu}(\partial_{\mu}+A_{\mu})\big)}{\Lambda^2}}\bigg)
\nn\\
&=&\int \frac{d^4p}{(2\pi)^4}\mathrm{Tr}\bigg(\langle y|p\rangle\langle p|\gamma_5
e^{-\frac{\big(\gamma_{\mu}(p_{\mu}-iA_{\mu})\big)^2}{\Lambda^2}}|x\rangle\bigg)_{\Lambda^2\rightarrow\infty, \ y\rightarrow x}
\nn\\
&=&
\lim_{\Lambda^2\rightarrow\infty}\bigg\lbrack\int d^4x\ \mathrm{Tr}\bigg(\frac{1}{8M^4}\gamma_5F_{\mu\nu}\gamma_{\mu}\gamma_{\nu}F_{\rho\sigma}\gamma_{\rho}\gamma_{\sigma}\bigg)
\nn\\
&&\times
\int\frac{d^4p}{(2\pi)^4}\ e^{-\frac{p^2}{\Lambda^2}}\bigg\rbrack
\nn\\
&=&
\frac{1}{16\pi^2}\int d^4x\ \epsilon_{\mu\nu\rho\sigma}\mathrm{Tr}(F_{\mu\nu}F_{\rho\sigma}),
\eea
where the field strength $F_{\mu\nu}$ is defined as
\bea
F_{\mu\nu}\equiv \partial_{\mu}A_{\nu}-\partial_{\nu}A_{\mu}+\lbrack A_{\mu}, A_{\nu}\rbrack.
\eea
We use the convention:
\bea
\epsilon_{0123}=1; \ \mathrm{Tr}(\gamma^5\gamma^{\mu}\gamma^{\nu}\gamma^{\rho}\gamma^{\sigma})=4\epsilon^{\mu\nu\rho\sigma}
\eea
and the following identity:
\bea
\gamma^{\mu}\gamma^{\nu}{\cal D}_{\mu}{\cal D}_{\nu}
&=&\frac{1}{2}\{\gamma_{\mu}, \gamma_{\nu}\}{\cal D}_{\mu}{\cal D}_{\nu}
+\frac{1}{2}\lbrack\gamma_{\mu}, \gamma_{\nu}\rbrack{\cal D}_{\mu}{\cal D}_{\nu}
\nn\\
&=&{\cal D}^2+\frac{1}{2}\gamma_{\mu}\gamma_{\nu}F_{\mu\nu},
\eea
where ${\cal D}_{\mu}$ is the covariant derivative
\bea
{\cal D}_{\mu}\equiv \partial_{\mu}+A_{\mu}.
\eea
The results show that $\mathrm{Tr}(\gamma_5)$ induces the topological charge.
Therefore, it reflects that the topological charge density induces the chiral anomaly (or the non-conservation of the current).

\subsection{Index Theorem on Lattice}
\noindent
When a Dirac matrix satisfies the Ginsparg-Wilson relation \cite{Ginsparg:1981bj}
\bea
\gamma_5D+\gamma_5=aD\gamma_5D,
\eea
the lattice artifact modifies the chiral transformation \cite{Luscher:1998pqa}:
\bea
\psi&\rightarrow&\exp\bigg\lbrack i\alpha\gamma_5\bigg(1-\frac{a}{2}D\bigg)\bigg\rbrack\psi; 
\nn\\
\bar{\psi}&\rightarrow&\bar{\psi}\exp\bigg\lbrack i\alpha\bigg(1-\frac{a}{2}D\bigg)\gamma_5\bigg\rbrack.
\eea
The chiral anomaly is also modified, and then the index theorem is realized through the expression \cite{Fujikawa:1979ay}:
\bea
\mathrm{Tr}\bigg\lbrack\gamma_5\bigg(1-\frac{a}{2}D\bigg)\bigg\rbrack
&=&\frac{1}{2}\sum_{\lambda}(v_{\lambda}, \gamma_5(2-aD)v_{\lambda})
\nn\\
&=&\frac{1}{2}\sum_{\lambda}(2-a\lambda)(v_{\lambda}, \gamma_5 v_{\lambda})
\nn\\
&=&n_+-n_-,
\nn\\
\eea
where $a$ is the lattice spacing, $v_{\lambda}$ is an eigenvector of $D$ associated with the eigenvalue $\lambda$, and $n_+$ and $n_-$ denote the number of left- and right-handed zero modes.
Only eigenvectors with real eigenvalues, 0 and $2/a$, have the non-vanishing chirality \cite{Fujikawa:1979ay}:
\bea
(v_{\lambda=0}, \gamma_5 v_{\lambda=0})\neq 0; \  (v_{\lambda=\frac{2}{a}}, \gamma_5 v_{\lambda=\frac{2}{a}})\neq 0.
\eea
The topological charge is significantly influenced only by the zero modes \cite{Fujikawa:1979ay}.
\\

\noindent
When a Dirac matrix satisfies
\bea
\gamma_5D+D\gamma_5=0,
\eea
the chiral anomaly is only contributed from $\mathrm{Tr}(\gamma_5)$.
When tracing over the Dirac indices first, it shows zero already.
On a finite lattice, there is no chiral anomaly.
However, it is difficult to say that the lattice fermion theory is inconsistent due to the loss of the anomaly on a finite lattice.
We will give one example, showing that the topologically-nontrivial background gauge fields always possess zero topological charge on a finite lattice \cite{Chiu:2001bg,Chiu:2001ja}.
This example teaches us that it is tricky from the infinite lattice size limit \cite{Chiu:2001bg,Chiu:2001ja}.
The loss of the chiral anomaly is likely due to the consideration of a finite lattice.
The physical observable can still be obtained from the lattice simulation \cite{Chiu:2001bg,Chiu:2001ja}.

\subsection{Zero Mode and Chiral Anomaly}
\noindent
The Dirac operator is defined as follows \cite{Chiu:2001bg,Chiu:2001ja}
\bea
D=\frac{1}{a}D_c\bigg(1+\frac{1}{2}D_c\bigg)^{-1},
\eea
where
\bea
\gamma_5D_c+D_c\gamma_5=0,
\eea
we can use
\bea
D^{-1}=a\bigg(\frac{1}{2}+D_c^{-1}\bigg)
\eea
to show the Ginsparg-Wilson relation or \cite{Chiu:2001bg,Chiu:2001ja}
\bea
\gamma_5D^{-1}+D^{-1}\gamma_5=a\gamma_5.
\eea
This Dirac matrix satisfies the $\gamma_5$-Hermiticity \cite{Chiu:2001bg,Chiu:2001ja}.
The zero mode of the Dirac matrix is also a zero mode of $D_c$ \cite{Chiu:2001bg,Chiu:2001ja}.
Suppose the spectrum of $D_c$ does not contain any poles for a topologically-nontrivial gauge background.
In that case, the Dirac matrix loses topological zero modes.
The solution with properties that are exponentially local, free of doublers, and exhibit expected continuum behavior was constructed in Ref. \cite{Chiu:2001bg}.
Due to the loss of the topological zero modes, the topological charge is zero even if the chiral anomaly appears \cite{Chiu:2001bg,Chiu:2001ja}.
The homogeneous part of the topological charge density (contributing to the zero topological charge) is recovered in the infinite lattice size and continuum limits as the expected result \cite{Chiu:2001bg,Chiu:2001ja}.
Hence, the winding part of the topological charge density (contributing to the non-zero topological charge) is suppressed when the lattice size approaches infinity \cite{Chiu:2001bg,Chiu:2001ja}.
To have the non-trivial contribution of the topological charge, we should carefully treat the infinite summation of the infinitesimal value. 
Because $\mathrm{Tr}(\gamma_5)=0$ loses the smooth continuum limit when the doublers disappear, taking the infinite lattice size limit is subtle \cite{Fujikawa:1999ku}. 
Due to similar issues with anomaly calculation, assessing consistency at a finite lattice level is challenging.

\section{Outlook and Future Directions}
\label{sec:7}
\noindent
We identified the challenges and trade-offs in putting chiral fermions on the lattice, particularly regarding chiral symmetry and Hermiticity.
The Nielsen-Ninomiya theorem constrains the possibilities for chiral fermions on a lattice \cite{Nielsen:1980rz,Nielsen:1981xu}.
Typical methods, like Wilson \cite{Wilson:1974sk} and overlap fermions \cite{Neuberger:1998wv}, circumvented the theorem at the cost of local action.
Non-Hermitian formulations stood out here as they sidestep the no-go theorem by abandoning Hermiticity \cite{Stamatescu:1993ga}, allowing for both chiral symmetry and local action—though this introduces other technical issues for simulation.
\\

\noindent
The usage of even-flavor fermions as a means to achieve a positive-definite partition function is a significant workaround \cite{Guo:2021sjp}.
It allows MC simulations in the non-Hermitian lattice formulation \cite{Guo:2021sjp}.
While the exact chiral symmetry in non-Hermitian setups challenges traditional views on the chiral anomaly on the lattice \cite{Guo:2021sjp}, this could still be resolved or recovered in the infinite-volume limit.
This area remains open and intriguing for further exploration, especially regarding how the anomaly behavior adjusts in the infinite-volume limit.
\\

\noindent
This area in finite-density QCD is complex yet essential, as the sign problem prevents standard Monte Carlo simulations at a real chemical potential.
Using imaginary chemical potential and analytical continuation is one of the few viable approaches, and combining it with a lattice formulation that respects exact chiral symmetry can provide a significant advantage.
Resummation methods can be potent here, allowing us to explore non-perturbative effects and, ideally, to verify results from lattice simulations in strong coupling regimes.
Matching these two methods offers robust insights into QCD's phase transitions and other phenomena.
\\

\noindent
Focusing on low-lying eigenmodes for non-Hermitian lattice QCD could streamline simulations while retaining physical insights, especially in the absence of zero modes.
This selective approach could reveal whether low-mode truncations approximate the continuum results within acceptable tolerances, particularly for observables sensitive to chiral symmetry breaking or confinement.
Establishing an efficient method is meaningful.
Because the Dirac matrix does not possess the zero modes \cite{Guo:2021sjp}, more consistent checks between the lattice simulation and the continuum field theory should confirm whether the topologically trivial configurations could reproduce physical results \cite{Chiu:2001bg,Chiu:2001ja}.

\section*{Acknowledgments}
\noindent
We want to express our gratitude to Sinya Aoki, Xing Huang, and Hersh Singh for their helpful discussion.
CTM would like to thank Nan-Peng Ma for his encouragement. 
%CTM acknowledges the Nuclear Physics Quantum Horizons program through the Early Career Award (Grant No. DE-SC0021892). 
HZ acknowledges the Guangdong Major Project of Basic and Applied Basic Research (Grant No. 2020B0301030008) and the National Natural Science Foundation of China (Grant No. 12105107).

%\appendix

\end{document}